\documentclass[twocolumn]{aastex63}

\usepackage{mathrsfs}
\usepackage[]{txfonts}
\usepackage{comment}

\shortauthors{Miyakawa et al.}
\shorttitle{Photometric Variations of Young Cool Stars}
\begin{document}

\title{Wavelength Dependence of Activity-Induced Photometric Variations for Young Cool Stars in Hyades}

\def\myemail{miyakawa@geo.titech.ac.jp}
\def\titech{Department of Earth and Planetary Sciences, Tokyo Institute of Technology, Meguro-ku, Tokyo, 152-8551, Japan}
\def\oao{Okayama Astrophysical Observatory, National Astronomical Observatory of Japan, Asakuchi, Okayama 719-0232, Japan}
\def\naoj{National Astronomical Observatory of Japan, NINS, 2-21-1 Osawa, Mitaka, Tokyo 181-8588, Japan}
\def\tokyo{Department of Astronomy, Graduate School of Science, The University of Tokyo, Hongo 7-3-1, Bunkyo-ku, Tokyo, 113-0033, Japan}
\def\komaba{Komaba Institute for Science, The University of Tokyo, 3-8-1 Komaba, Meguro, Tokyo 153-8902, Japan}
\def\Canaria{Instituto de Astrof\'isica de Canarias, V\'ia L\'actea s/n, E-38205 La Laguna, Tenerife, Spain}
\def\abc{Astrobiology Center,  2-21-1 Osawa, Mitaka, Tokyo 181-8588, Japan}
\def\hawaii{Department of Earth Sciences, University of Hawaii at M$\tilde{a}$noa, Honolulu, HI 96822, USA}
\def\norcal{Department of Physics and Astronomy, The University of North Carolina at Chapel Hill, Chapel Hill, NC 27599, USA}

\author{Kohei Miyakawa}\affiliation{\titech}
\author[0000-0003-3618-7535]{Teruyuki Hirano}\affiliation{\abc}\affiliation{\naoj}
\author[0000-0002-4909-5763]{Akihiko Fukui}\affiliation{\komaba}\affiliation{\Canaria}
\author{Andrew W. Mann}\affiliation{\norcal}
\author[0000-0002-5258-6846]{Eric Gaidos}\affiliation{\hawaii}
\author{Bun'ei Sato}\affiliation{\titech}
\begin{abstract}
	We investigate photometric variations due to stellar activity 
	which induce systematic radial-velocity errors (so-called ``jitter") 
	for the four targets in the Hyades open cluster observed by the $K2$ mission 
	(EPIC 210721261, EPIC 210923016, EPIC 247122957, and EPIC 247783757).
	Applying Gaussian process regressions to the $K2$ light curves 
	and the near-infrared (NIR) light curves observed with the IRSF 1.4-m telescope,
	we derive the wavelength dependences of the photometric signals due to stellar activity.
	To estimate the temporal variations in the photometric variability amplitudes 
	between the two observation periods of $K2$ and IRSF, separated by more than 2 years,
	we analyze a number of $K2$ targets in Hyades that have also been observed in Campaigns 4 and 13
	and find a representative variation rate over 2 years of $38\%\pm71\%$.
	Taking this temporal variation into account, 
	we constrain projected sizes and temperature contrast properties of the starspots in the stellar photosphere
	to be approximately $10\%$ and 0.95, respectively.
	These starspot properties can induce relatively large differences in the variability amplitude over different observational passbands, 
	and we find that radial-velocity jitter may be more suppressed in the NIR than previously expected.
	Our result supports profits of on-going exoplanet search projects that are attempting to detect or confirm young planets in open clusters via radial-velocity measurements in the NIR. 
\end{abstract}
\section{Introduction}\label{sec:intro}
	Exoplanetary studies have made great progress 
	with the help of space telescope missions such as NASA's $Kepler$ space telescope \citep{Borucki2010}.
	In particular, $Kepler$'s secondary mission, $K2$, conducted a systematic survey
	of young transiting planets in stellar clusters ($< 1$ Gyr), that were not included in the original mission 
	(e.g., Hyades \citep[650 Myr,][]{Martin2018},
	Pleades \citep[112 Myr,][]{Dahm2015}, and Upper Scorpius \citep[11 Myr,][]{Pecaut2012}).
	Several planets have been confirmed and/or validated around these young stars \citep[e.g.,][]{Mann2016,Mann2016b,Mann2017}, 
	which are important to study to determine the formation and evolutionary processes of exoplanets, 
	as well as their primordial atmospheres.
	More recently, an all-sky survey by the $TESS$ mission \citep{Ricker2015} also revealed 
	young exoplanets in other stellar associations \citep[e.g.,][]{Newton2019, Rizzuto2020, Mann2020}. 

	Even though the sky regions being surveyed have expanded,
	the number of planets detected around young stars remains far more limited ($\sim 30$)
	than those around their older counterparts.
	One reason for the small number of detections are the high surface activities of these young 
	host stars.
	When the brightness of the stellar surface is inhomogenious (e.g., includes spots and plages),
	apparent modulations due to stellar rotation appear in radial velocity (RV) measurements.
	These activity-induced apparent signals, so-called ``RV jitters", 
	prevent the detection of true planetary signals \citep[e.g.,][]{Queloz2001,Paulson2004}.
	In particular, less is known concerning the properties of RV jitter for low-mass stars 
	(M dwarfs) as a result of their intrinsic faintness, 
	despite the fact that starspots tend to exist longer on M dwarfs than on solar-type 
	stars and could 
	affect their long-term RV variability \citep{Robertson2014, Robertson2020, Davenport2015}. 
	Recent studies have suggested that 
	planetary radii around young stars are significantly larger than those around
	old stars in the low-mass region \citep[$< \,0.6 \,M_\odot$, e.g.,][]{Obermeier2016, Mann2018, Rizzuto2020}.
	Studies of young low-mass stars are important to 
	test for the scenarios that explain the difference in the radius distributions, 
	including atmospheric escape \citep[e.g.,][]{Owen2019}.
	
	In general, it is known that RV jitter is reduced in the near-infrared (NIR)
	relative to optical wavelengths 
	because the contrast between starspots and the photosphere is smaller at longer wavelengths \citep{Bean2010, Reiners2010, Anglada-Escude2013, Tal-Or2018, Robertson2020}.
	However, the number of observational samples available for systematic studies of stellar activity 
	is still small and the detailed properties of starspots are concealed, especially for young M dwarfs.
	Measuring stellar RVs is one of the promising methods to study the detailed properties of jitter, 
	such as their wavelength dependence \citep[e.g.,][]{Robertson2020}; 
	however, it is time-consuming to obtain a sufficiently large number of observations, 
	especially for optically faint targets such as M dwarfs.
	\citet{Frasca2009} have approached this problem using photometric observations; 
	however, their targets have been limited to pre-main-sequence stars ($\sim10$ Myr)
	that show large photometric variations ($>10\,\%$).
	Therefore, reduced jitter in the NIR and its wavelength dependence have not been robustly established for various types of stars of different ages.
	For a more accurate understanding of young exoplanets, it is necessary to 
	constrain the behavior of stellar rotational activity using both photometry and spectroscopy.

 	In this study, we evaluate the observational behavior of stellar rotational activity 
 	in the NIR.
	We focus on M dwarfs in the Hyades open cluster 
	\citep[$650 $ Myr; ][]{Martin2018} that were photometrically observed by the $K2$ mission,
	and we investigate stellar jitter using multicolor photometry combining the $K2$ and NIR light curves.
	We also estimate the starspot sizes and temperatures using a toy model.
	Our approach allows us to roughly understand the properties of starspots for targets 
	showing relatively small ($\sim1\,\%$) simple photometric variabilities with 1-m-class telescopes.
 
	This paper is organized as follows.
	In Section \ref{sec:observation}, 
	we introduce our targets and follow-up observations using the NIR multicolor photometry. 
	We show the analytic procedure for the targets involving Gaussian process regression in Section \ref{sec:GP}. 
	Section \ref{sec:spots} presents an estimation of the starspot properties.
	Finally, we discuss our interpretations of the results compared to previous studies 
	and give a conclusion in Section \ref{sec:discussion}.

	\begin{table*}[t]
	\centering
	\caption{\small Stellar properties of our targets.}\label{tab:stellarparam}
  		\begin{tabular}{ccccc} \hline
    		\hline
     		& EPIC 210721261 & EPIC 210923016 & EPIC 247122957 & EPIC 247783757\\ 
    		\hline \\
		Measured Property \\
		Apparent $B$ magnitude (1)	& $14.75 \pm0.05$ 	& $15.52\pm0.12$
			& $15.07\pm0.04$	& $15.55\pm0.04$\\
		Apparent $V$ magnitude (1)	& $13.31\pm0.05$ 	& $14.01\pm0.10$
			& $13.58\pm0.02$	& $14.05\pm0.05$\\
		Apparent $K$ magnitude (1) 	& $8.69\pm0.02$ 	& $9.22\pm0.02$
			& $8.92\pm0.02$	& $9.32\pm0.02$\\
		Parallax [mas] (2)	& $23.14\pm0.05$	& $20.90\pm0.07$	
		& $19.77\pm0.13$	&	$20.45\pm0.07$\\
		Astrometric Goodness of Fit: {\tt GOF\_AL} (2)	& 14.9 	& 21.0	&	51.0		& 26.2\\
		Astrometric Excess Noise Significance: {\tt D} (2)	&  0.0	& 11.3 	& 	48.2		& 18.5 \\ 
		Renormalised Unit Weight Error: {\tt RUWE} (2) 	& 1.21	& 1.18 	& 	6.51 	& 1.06 \\ 
		$BP - RP$ color (2)	& 2.39 &2.54 &2.47&2.53\\
		\\
			
   		Derived Property \\
		Rotation period : $P_{\rm rot}$ (3)& $1.5409\pm0.0001$ & $1.1570\pm0.0005$ 
			& $1.5457\pm0.0002$	& $1.8276\pm0.0002$\\
    		Effective temperature : $T_{\rm eff}$ [K] (4)	& $3488\pm79$	 &$3428\pm86$
			& $3462\pm76$	& $3443\pm78$\\
		Radius : $R_*$ [$R_\odot$] (5)	& $0.52\pm0.02$	 &$0.47\pm0.02$
			& $0.53\pm0.02$	& $0.47\pm0.01$\\
		Mass : $M_\star$ [$M_\odot$] (6)	& $0.52\pm0.01$	 &$0.47\pm0.01$
			& $0.53\pm0.01$	& $0.47\pm0.01$\\
		Surface gravity : $\log{g}$ (5)(6) & $4.72\pm0.02$ & $4.76\pm0.03$
			& $4.71\pm0.02$& $4.77\pm0.03$\\ \\
		{\it Effective temperature} : $T_{\rm eff}$ [K] (7)	& $3624\pm45$	 &$3532\pm45$
			& $3575\pm45$	& $3539\pm45$ \\
		{\it Radius} : $R_*$ [$R_\odot$] (8)	& $0.43\pm0.05$	 &$0.40\pm0.04$
			& $0.42\pm0.05$	& $0.41\pm0.05$
			\\ \\ \hline
		\end{tabular}
		\begin{flushleft}
  			{\bf References :} (1):  VizieR database (https://vizier.u-strasbg.fr/viz-bin/VizieR), 
			(2): Gaia database (https://gea.esac.esa.int/archive), 
			(3): This study, (4): $V-J$ vs $T_{\rm eff}$ in \citet{Mann2015},
			(5): $M_{K}$ vs $R_{*}$ in \citet{Mann2015},
			(6): $M_{K}$ vs $M_{\star}$ in \citet{Mann2019},
			(7): $BP-RP$, [Fe/H] vs $T_{\rm eff}$ in \citet{Mann2015}, 
			and (8): $T_{\rm eff}$ vs $R_{*}$ in \citet{Mann2015}.
  	 	\end{flushleft}
	\end{table*}

\section{Targets and Observations}\label{sec:observation}
	\begin{figure}[]
 		\centering
 		\includegraphics[width=8cm]{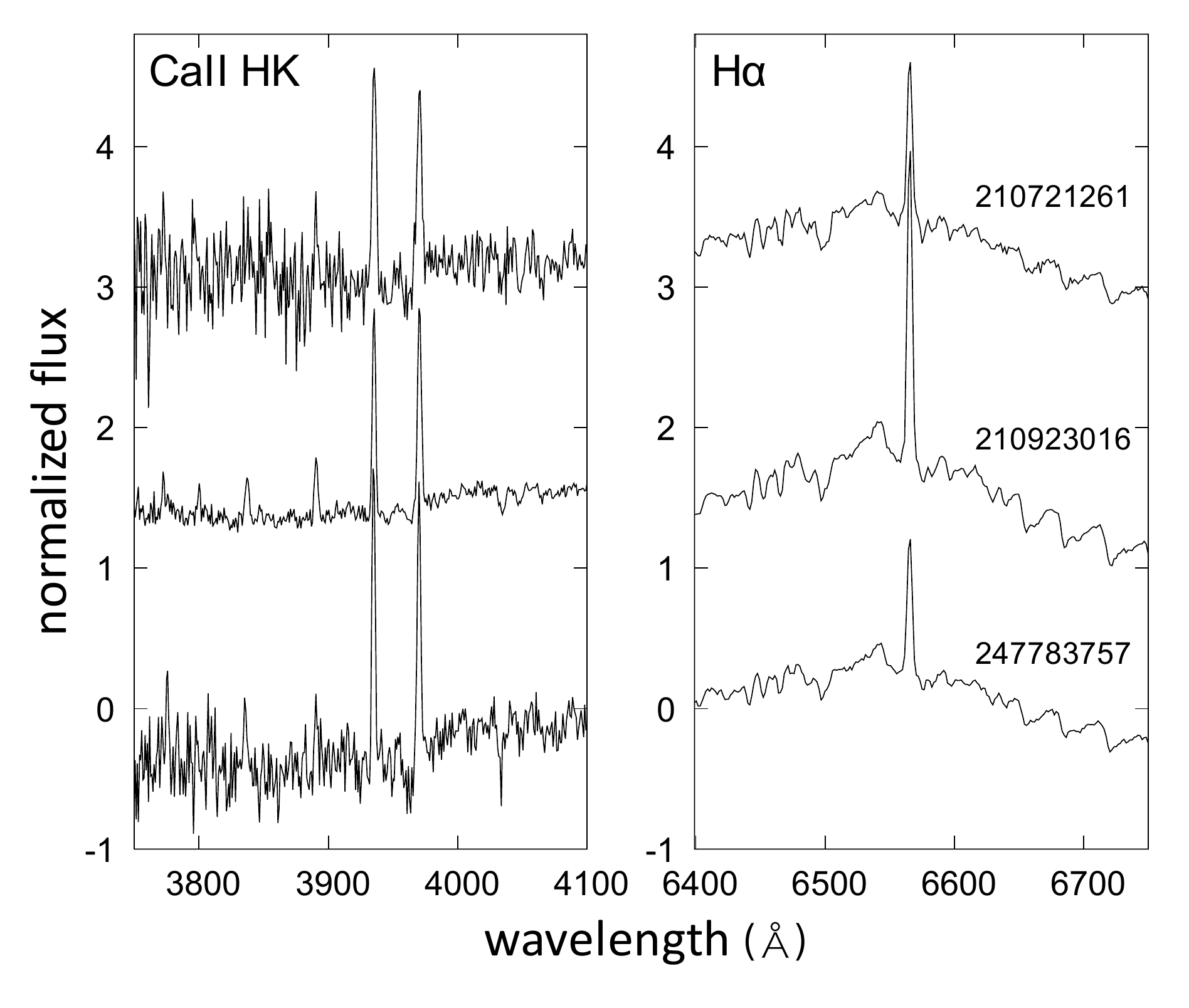}
 		\caption{\small LAMOST low-resolution spectra for the three targets. 
		The left and right panels show features around CaII HK and $\rm H\alpha$ emission lines, respectively.}
 		\label{pic:spectra}
	\end{figure}
	\begin{figure}[]
 		\centering
 		\includegraphics[width=8cm]{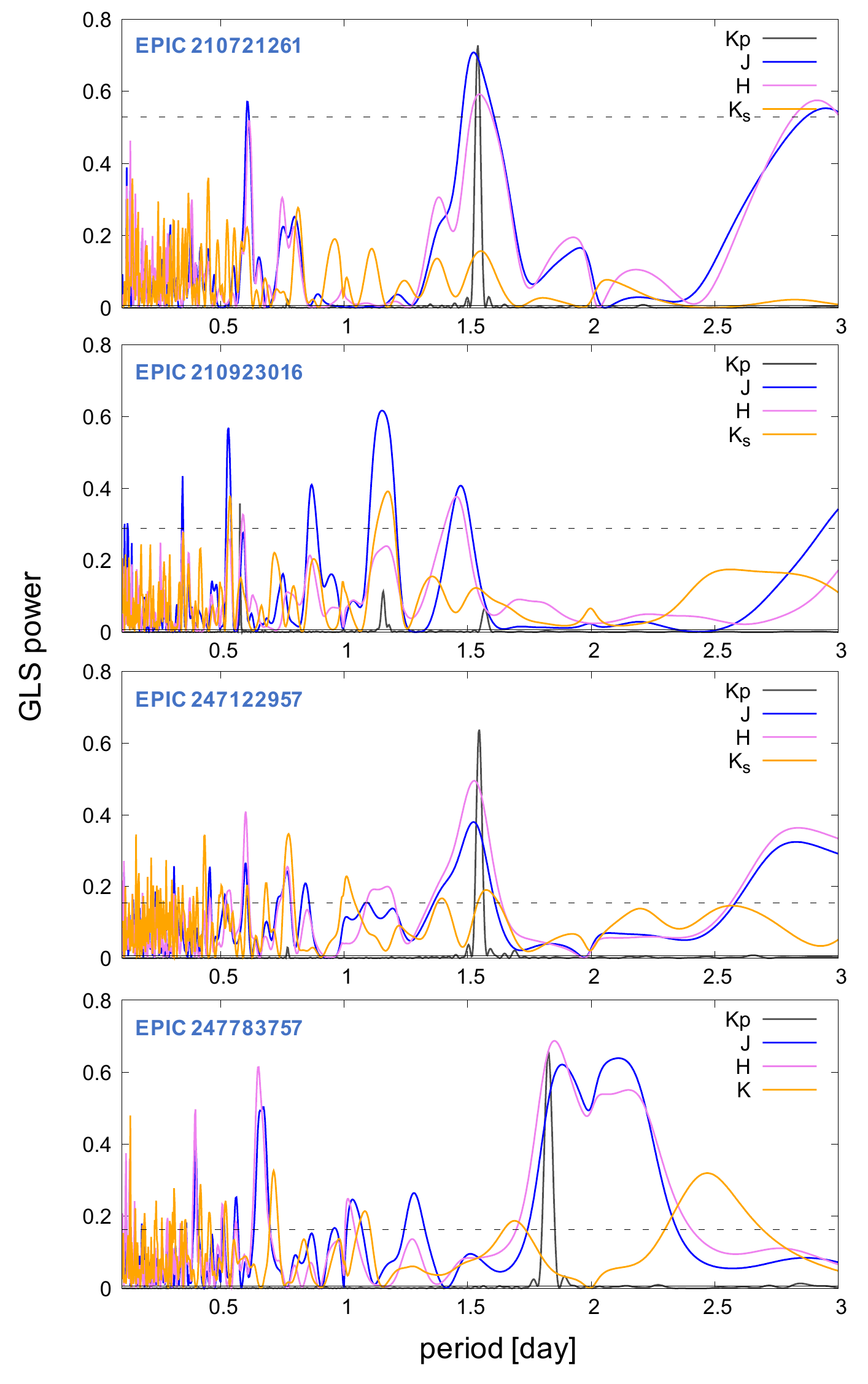}
 		\caption{\small Generalized Lomb-Scargle periodograms for our four targets.
		Each solid colored line represents an observational passband.
		The dashed horizontal line represents 0.1\% of the false alarm probability (FAP) in the $J$ band.
		The solid horizontal line represents the 0.1\% FAP in the $Kp$ band.}%
 		\label{pic:lombscargle}
	\end{figure}
	\subsection{K2 Targets in the Hyades Open Cluster}\label{sec:K2}
	$K2$ is the secondary mission of the $Kepler$ space telescope 
	to detect transiting exoplanet candidates along the ecliptic plane \citep{Howell2014}.
	Some $K2$ campaign fields include young clusters. 
	To reduce systematic effects when studying the properties of starspot, 
	such as those associated with age and metallicity, 
	we focused on young stars belonging to a single open cluster. 
	We selected targets from the Hyades open cluster 
	with typical metallicity [Fe/H] and age values of approximately 
	$0.14\pm0.05$ and $650\pm70$ Myr, respectively \citep{Perryman1998, Brandt2015, Martin2018}.
	
	We picked four preferable targets for ground-based follow-up observations
	that are relatively bright and have short rotation periods ($< 10 {\rm mag}_{K}$ and $< 3$ days)
	from the VizieR table \footnote{https://vizier.u-strasbg.fr/viz-bin/VizieR-3?-source=J/ApJ/879/100/table3} 
	reported in \citet{Douglas2019}: 
	EPIC 210721261, EPIC 210923016, EPIC 247122957, and EPIC 247783757.
	Transiting planetary candidates have not been detected for these targets.
	The $K2$ data were collected on Campaign 13,  which was from March 8 to May 27, 2017.
	We use the Pre-search Data Conditioned-Simple Aperture Photometry light curves dowloaded 
	from the Mikulski Archive for Space Telescopes 
	portal\footnote{http://archive.stsci.edu/k2/epic/search.php}.
	For this study, 
	we normalized the overall long-term systematic modulations 
	using a fifth-order polynomial function and removed flux outliers via $5-\sigma$ clipping,
	to focus on periodic variations due to stellar rotation.
	We summarize the properties of the targets in Table \ref{tab:stellarparam}.
	The stellar effective temperatures and radii are derived by $V-J$ vs $T_{\rm eff}$ relation and
	$M_{K}$ vs $R_*$ relation from \citet{Mann2015}, respectively;
	the masses are by $M_{K}$ vs $M_{\star}$ relation from \citet{Mann2019} with the metallicity of 0.15.
	
	We also estimated the stellar properties using different equations, 
	because young active stars have large uncertainties in the colors due to their inhomogeneous surfaces
	\citep{Stauffer2003}.
	Additional estimations for $T_{\rm eff}$ and $R_{*}$ 
	using the $BP - RP$ vs $T_{\rm eff}$ and $T_{\rm eff}$ vs $R_{*}$ relations, respectively 
	from \citep{Mann2015}, 
	are listed in the bottom of Table \ref{tab:stellarparam}.
	These values deviate by about $1\sigma$ from the original estimations, 
	which means that there are large uncertainties in the broad-band colors.
	In addition, we checked the CaII HK (393.4 nm and 396.8 nm) 
	and ${\rm H\alpha}$ (656.3 nm) emission lines as activity indicators
	using low-resolution spectra observed with the the Large Sky Area Multi-Object Fiber Spectroscopic
	Telescope \citep[LAMOST; ][]{Cui2012}\footnote{http://dr6.lamost.org/v2/search}.
	The spectral data are available for EPIC 210721261, EPIC 210923016, and EPIC 247783757.
	All the three targets exhibit significant emissions at those lines as in Figure \ref{pic:spectra}.
	Thus, we conclude that our targets are very active  
	and it is possible that the systematic errors in the effective temperatures are underestimated.
	In later discussions, we will take into account this point to interpret the results more accurately.

	\subsubsection{Rotation Periods}
		For the selected target stars, 
		we first investigated the periodic modulations
		in the $K2$ light curves and determined the rotation period for each target. 
		To do so, 
		we applied the generalized Lomb-Scargle periodogram \citep[GLS;][]{Zechmeister2009}
		to unbinned K2 light curves;
		the results are shown in Figure \ref{pic:lombscargle} with the black solid line.
		In addition, we computed the autocorrelation function \citep[ACF; ][]{McQuillan2013}
		using the same datasets 
		to confirm the periods identified in the GLS periodogram.
		For EPIC 210721261, EPIC 247122957 and EPIC 247783757,
		the same periods were detected by both GLS and ACF 
		and we adopted these periods for the subsequent analyses.
		For EPIC 210923016, the highest peak was detected at 0.58 days in GLS.
		ACF, however, showed the highest peak at 1.16 days, which is twice the GLS period.
		The 0.58-day peak is an upper harmonic of 1.16 day and 
		likely due to multiple starspots on the surface,
		and we concluded that 1.16 days is the true rotation period of the star.
		Finally, via visual inspection, we confirmed semi-coherent modulations with the determined periods in the $K2$ light curves for all targets (Figure \ref{K2LC}).
	
	\subsubsection{Possibility of Binary}
		Because we are trying to constrain the starspot properties (e.g., sizes and
		temperatures) from multicolor photometric observations, it is important to 
		rule out the presence of nearby (companion) stars in the photometric aperture
		because, when a light curve is diluted by flux contamination from nearby stars, 
		the interpretation of the amplitude of the light curve modulations is more complicated. 
		To ensure of the absence of possible companion stars, 
		we inspected nearby stars listed in the 2MASS \citep{Cutri2003}, 
		SDSS \citep{Adelman-McCarthy2009}, and Gaia \citep{Gaia2018} catalogs
		on the VizieR website\footnote{http://vizier.u-strasbg.fr/viz-bin/VizieR},
		and found that there are no resolved companions within $3''$.
		However, it is difficult to completely eliminate the possibility of binaries,
		because high-precision adaptive optic observations have not been performed.
		Therefore, to evaluate the binarity of the targets, 
		we employed the thresholds described in \citet{Evans2018} 
		for the Astrometric Goodness of Fit in the Along-Scan direction ($\tt GOF\_AL$) 
		and the Significance of the Astrometric Excess Noise ($\tt D$) in the Gaia second data release \citep[DR2; ][]{Gaia2016, Gaia2018}.
		These parameters characterize the agreement 
		between an astrometric model and the data depending on the presence of unresolved companions.
		\citet{Evans2018} set the threshold of binarity condition to ${\tt GOF\_AL} > 20$ and ${\tt D} > 5$.
		In addition, we referred the Renormalised Unit Weight Error ($\tt RUWE$) statistics 
		in the Gaia early third data release \citep[EDR3; ][]{Gaia2020}, which is discussed in \citet{Stassun2021}.
		They suggest that $\tt RUWE$ is strongly correlated to binarity condition 
		for $1.0 < {\tt RUWE} < 1.4$.
		We list the $\tt GOF\_AL$ and $\tt D$ values in DR2 and $\tt RUWE$ values in EDR3 
		for our targets in Table \ref{tab:stellarparam}.
		
		For EPIC 210721261, the $\tt GOF\_AL$ and $\tt D$ values are 14.9 and 0.0, respectively, which are significantly lower than the thresholds;
		therefore, the possibility of a binary is low. 
		EPIC 247122957, whose $\tt GOF\_AL$ and $\tt D$ values are relatively large (51.0 and 48.2, respectively) 
		may host a companion.
		For the other two targets, EPIC 210923016 and EPIC 247783757,
		even though their $\tt GOF\_AL$ and $\tt D$ are slightly above the thresholds,
		we cannot confidently suggest their status as binaries.
		While the $\tt GOF\_AL$ and $\tt D$ values were recently updated in EDR3 \citep{Gaia2020},
		we cannot directly compare them, 
		because the thresholds in \citet{Evans2018} were derived from the DR2 data.
		However, we found that the EDR3 $\tt GOF\_AL$ values were all well under 20, 
		except that of EPIC 247122957.
		With regard to $\tt RUWE$, the value for EPIC 247122957 is high and deviated 
		from the correlated range (6.51),
		while these for the other targets are relatively small ($\leq 1.2$).
		Consequently, EPIC 210721261, EPIC 210923016, and EPIC 247783757 seem to be single stars
		with high probability from the aspect of astrometry.
		
		\citet{Douglas2019} mentioned  that rapid rotating early M-dwarfs in Hyades
		with periods of $\sim$ 1 day are likely binaries,
		whereas the typical period of them is $\sim $10 days.
		To discuss the binarities further, we also derived stellar radii based on different photometry.
		We show the additional radii estimated with $T_{\rm eff}$ vs $R_{*}$ relation in \citet{Mann2015}  
		to Table \ref{tab:stellarparam} in italic;
		the $T_{\rm eff}$ is derived from $V-J$ color.
		There are approximately $2 -\sigma$ deviations from the radii derived from $M_{K}$ for all the targets.
		One explanation for these systematics is that our targets may be entirely binaries.
		If it was true, fluxes from the companions are likely not dominant,
		because the periodograms in $Kp$ band show single peaks excluding harmonics for all the targets.
		In Section \ref{sec:discussion}, we test a dilution effect by a companion star
		to discuss uncertainties in the case where the targets are binaries.
	\begin{figure}[]
 		\centering
 		\includegraphics[width=8cm]{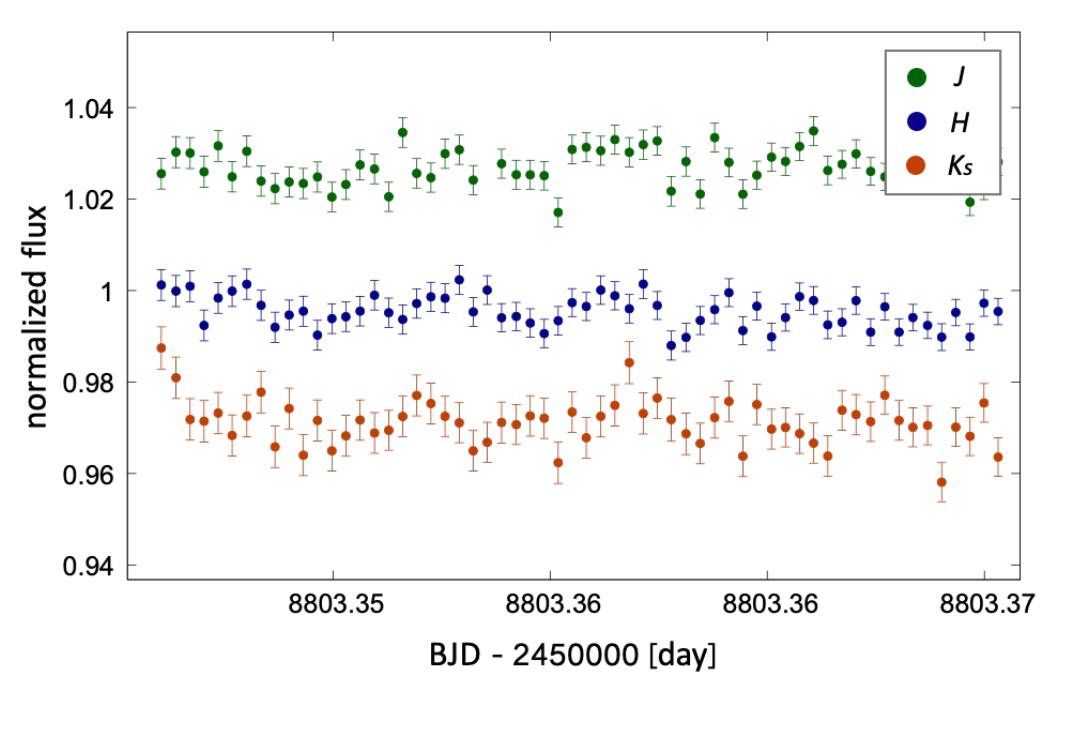}
 		\caption{\small  Example of a light curve segment observed with IRSF. 
			The light curve is for EPIC 210721261  
			and was extracted with the customized pipeline described in \citet{Fukui2011}.}%
 		\label{pic:local_obs}
	\end{figure}
	\begin{figure*}[t]
 		\centering
 		\includegraphics[width=16cm]{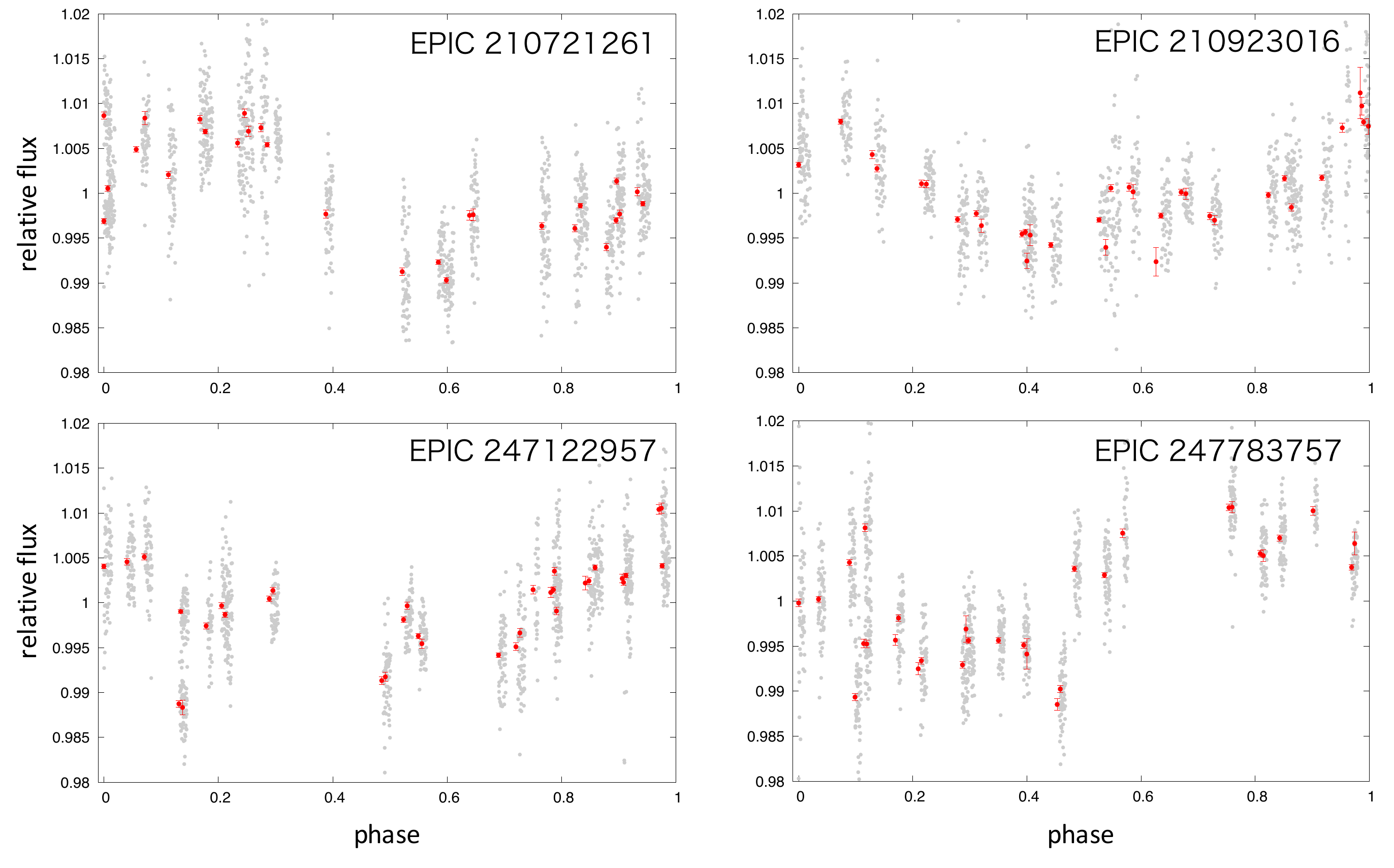}
 		\caption{\small Folded light curves in $J$ band observed with IRSF/SIRIUS. The gray and red points are unbinned and binned light curves, respectively.}%
 		\label{pic:irsf}
	\end{figure*}
	\subsection{Follow-up Observations, 
		IRSF 1.4m / SIRIUS} \label{sec:targetIRSF}
		From November 14 to 26, 2019, we conducted follow-up observations with 
		the Simultaneous Infrared Imager for Unbiased Survey \citep[SIRIUS;][]{Nagashima1999}
		on the IRSF 1.4 m telescope at the South Africa Astronomical Observatory.
		SIRIUS is equipped with three 1 k $\times$ 1 k HgCdTe detectors 
		with a pixel scale of $0\farcs45\,{\rm pixel}^{-1}$. 
		This enables us to take three NIR images in the $J$, $H$, and $K_s$ bands 
		simultaneously with the same exposure time for all bands.
		Setting the exposure times to 30 s with a dead time of approximately 8 s for all bands,
		we observed the 4 targets in rotation during a night. 
		Each target was observed for approximately 30 minutes per visit, 
		and was visited two or three times per night.
		The weather conditions were generally good,
		and observations were carried out nearly every night.
		
		Aperture photometry for each raw fits file has been performed using the customized pipeline 
		described in \citet{Fukui2011}.
		For each target, we employed two or three stars in the same frame as reference stars,
		and checked that their flux variations were negligible.
		{ The flux error $\sigma$ for each data point was calculated 
		as in Section 2.2 of \citet{Fukui2011}.
		We show an example of extracted light curves for EPIC 210721261 in Figure \ref{pic:local_obs}.}
		Because there are systematic errors due to the observational circumstances, 
		we applied systematic correction to the light curves, as described in \citet{Fukui2013};
		here, we employed the pixel positions of the target and airmass as correction parameters.
		Then, we binned the light curves into 0.05-day sized bins adopting the weighted mean for each bin.
		{The weighted mean and propagated error values in each bin were derived 
		as $x_{\rm wm} = \sum{\omega_i x_i}/\sum{\omega_i}$ 
		and $\sigma_{\rm wm} = 1/\sqrt{\sum \omega_{i}}$, respectively,
		where the weight $\omega_{i}$ was given by $1/{\sigma_{i}^2}$.
		Note that 
		we only consider errors in the aperture photometry, 
		and do not take into account other factors such as instrumental noise and/or weather conditions,
		which may possibly lead to underestimated total errors.
		It is difficult to evaluate these factors directly, 
		because photometry was performed in a discrete manner over the course of two weeks.
		In Section \ref{ss:jointGP}, we will explain how to treat these uncertainties 
		to estimate astrophysical signals more accurately.}
		
		We tested the reproducibility of the periodicities identified in the $K2$ photometry 
		using the GLS periodograms.
		The results are shown in Figure \ref{pic:lombscargle} as solid colored solid lines for all four targets.
		The horizontal dashed line represents  the false alarm probability of 0.1\% 
		\citep[FAP; ][]{Zechmeister2009} in the $J$ band.
		For almost targets, peaks higher than the 0.1\% $\mathrm{FAP}$ lines  
		were detected in the $J$ and $H$ bands,
		and the periodicities in the $K_s$ band were not detected with sufficient significance.
		This is because the detector is known to be unstable,
		and the systematic errors were larger than the astrophysical signals for the $K_s$ band.
		{ Overall, these periodogram results ensure the accuracies of our observational data
		and their reductions, 
		although their uncertainties are likely underestimated.}
		We show the light curves for the four targets in the $J$ band observed 
		with IRSF/SIRIUS in Figure \ref{pic:irsf}.
		The light curves were folded with the periods detected using the GLS.

\section{GP Regression to the Light Curves}\label{sec:GP}

	We applied Gaussian process (GP) regression 
	to quantify the behaviors of starspots in the light curves 
	\citep{Rasmussen2006, Haywood2014, Grunblatt2015, Hirano2016}.
	Detail of GP is described in Appendix \ref{apa}.
	In the following, we explain how to treat the observed light curves with GP.

	\subsection{Analysis of the $K2$ Light Curves}\label{GP_K2}
	\begin{figure*}[t]
 	\centering
 	\includegraphics[width=18cm]{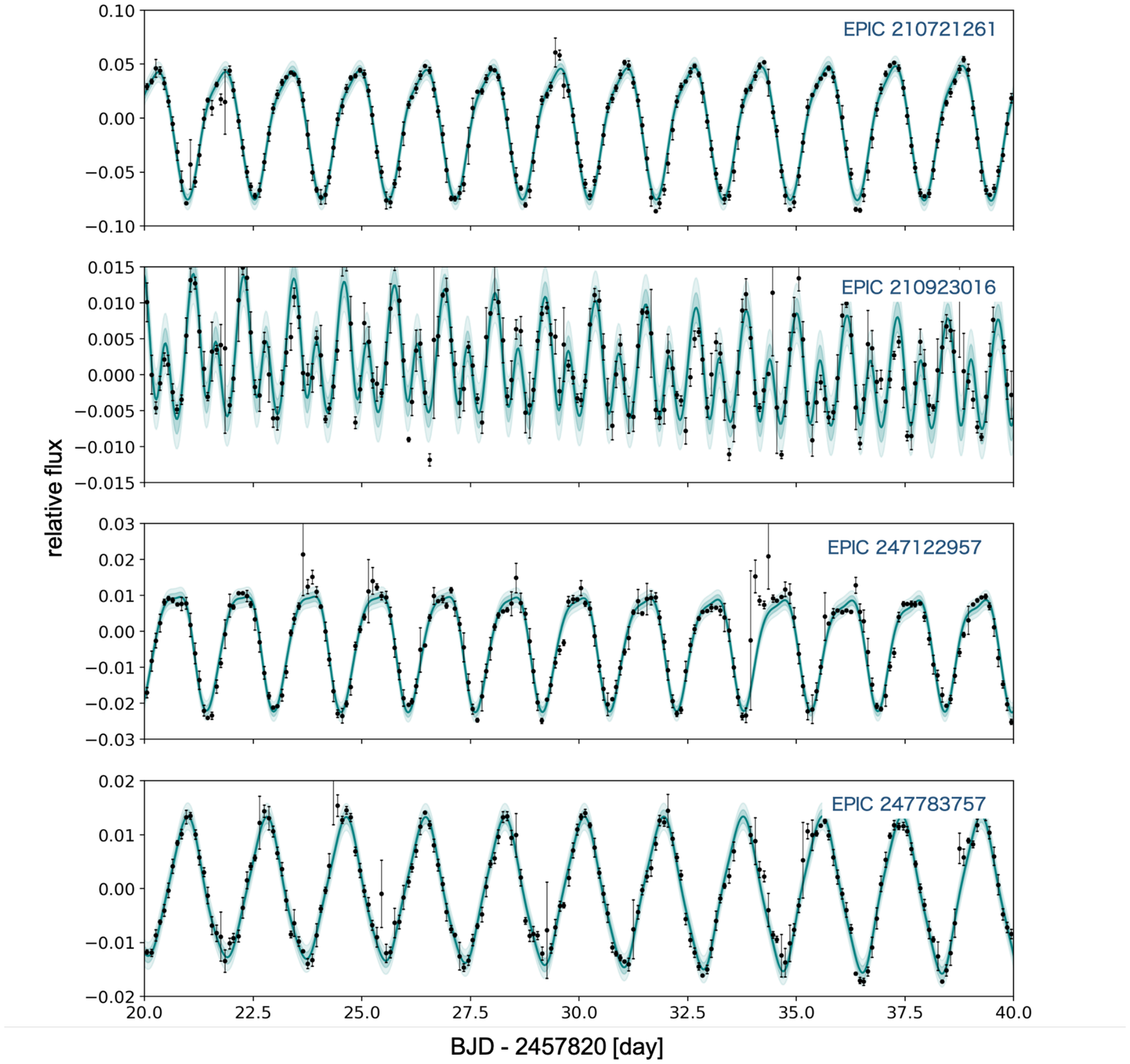}\\
 	\caption{\small 
	{Examples of the $K2$ light curves for our targets binned with 0.1 day.
	The binned values and error bars were calculated as the mean and standard deviation 
	weighted by the photon noise for each bin, respectively.}
	The blue solid line and regions represent the result of GP analyses.}%
 	\label{K2LC}
	\end{figure*}
	
	\begin{table*}[t]
	\centering
	\caption{\small Gaussian process (GP) hyperparameters}\label{tab:1}
  		\begin{tabular}{crrrr} \hline
    		\hline
     		& EPIC 210721261 & EPIC 210923016 & EPIC 247122957 & EPIC 247783757\\
    		    		\hline
    		Hyperparameters &\\
    		(optimized in Section \ref{GP_K2})\\ \\
    		$h$  	& $0.085$ 		& $0.012$ &
    			$0.018$ 		& $0.014$\\
		$\theta$	& $1.5409$ & 	$1.1587$ &
			$1.5475$	&	$1.8286$
			\\
    		$w$ 	& $1.22$	& $0.67$ 	&
    			$0.86$		& $0.94$\\
    		$\lambda$	& $51.51$		& $22.49$	&
    			$24.21$			& $25.32$	\\ 
    		$\sigma$ 	& $0.0046$ & $0.0020$ 	&
    			$0.0013$ 	& $0.0012$\\ \\
    		\hline \\
		(estimated by the joint analysis in Section \ref{ss:jointGP})
		\\ \\
    		$h_{Kp}$ 	& $0.0802_{-0.0127}^{+0.0167}$	&	$0.0126_{-0.0015}^{+0.0020}$ &
    			$0.0213_{-0.0025}^{+0.0030}$		&	$0.0215_{-0.0030}^{+0.0037}$\\ 
    		$\sigma_{Kp}$ 	& $0.0046_{-0.0002}^{+0.0002}$	&	${0.0020}_{-0.0001}^{+0.0001}$ &
    			$0.0013_{-0.0001}^{+0.0001}$		&	$0.0012_{-0.0001}^{+0.0001}$\\ \\ \\
    		$h_{J}$	& $0.0135_{-0.0048}^{+0.0082}$	& 	$0.0052_{-0.0018}^{+0.0027}$ &
    			$0.0054_{-0.0024}^{+0.0036}$		&	$0.0120_{-0.0041}^{+0.0071}$\\ 
    		$\sigma_{J}$	& $0.0003_{-0.0002}^{+0.0003}$	&	$0.0006_{-0.0003}^{+0.0005}$ &
    			$0.0016_{-0.0006}^{+0.0009}$		&	$0.0006_{-0.0004}^{+0.0005}$\\ \\
		$h_{sq, J}$		&	\multicolumn{4}{c}{$0.0032_{-0.0004}^{+0.0004}$}\\
		${\lambda}_{sq, J}$	&	\multicolumn{4}{c}{$0.0828_{-0.0213}^{+0.0255}$}\\ \\
			\\
    		$h_{H}$	& $0.0116_{-0.0047}^{+0.0097}$ 	& 	$0.0044_{-0.0024}^{+0.0035}$ &
    			$0.0055_{-0.0021}^{+0.0032}$		&	$0.0089_{-0.0028}^{+0.0049}$\\
    		$\sigma_{H}$	& $0.0005_{-0.0003}^{+0.0004}$	&	$0.0031_{-0.0006}^{+0.0008}$ &
    			$0.0011_{-0.0005}^{+0.0006}$		&	$0.0010_{-0.0004}^{+0.0005}$\\ \\
		$h_{sq, H}$		&	\multicolumn{4}{c}{$0.0032_{-0.0004}^{+0.0005}$}\\
		${\lambda}_{sq, H}$&	\multicolumn{4}{c}{$0.0700_{-0.0273}^{+0.0370}$}\\ \\
			\\
    		$h_{K_s}$	& $0.0094_{-0.0064}^{+0.0104}$	& 	$0.0058_{-0.0034}^{+0.0048}$ &
    			$0.0113_{-0.0063}^{+0.0102}$		&	$0.0049_{-0.0033}^{+0.0063}$\\
    		$\sigma_{K_s}$	& $0.0012_{-0.0008}^{+0.0015}$	& 	$0.0007_{-0.0005}^{+0.0018}$ &
    			$0.0010_{-0.0007}^{+0.0030}$		&	$0.0026_{-0.0012}^{+0.0030}$\\ \\
    		$h_{sq, K_s}$		&	\multicolumn{4}{c}{$0.0091_{-0.0011}^{+0.0010}$}\\
		${\lambda}_{sq, K_s}$	&	\multicolumn{4}{c}{$0.0635_{-0.0133}^{+0.0249}$}
		\\ \\
    		\hline

  		\end{tabular}
	\end{table*}

	First, we analyzed the $K2$ light curves using GP 
	to reproduce the flux variations due to stellar rotation, 
	because these curves were sufficiently precise and collected over a long period with good cadence.
	We binned the light curves into 0.1-day ranges
	considering the computational cost and the rotation period.
	We employed the quasi-periodic kernel (${\bf K}_{qp}$; Equation (\ref{kernel_qp})) in the GP regressions.
	This is because signals induced by stellar jitter show both periodicities due to stellar rotations 
	and coherent variations due to surface activities.
	To optimize the hyperparameters, 
	we used the Marcov Chain Monte Carlo (MCMC) method \citep{Foreman-Mackey2013} 
	and added Gaussian priors based on the rotation periods in Table \ref{tab:1} for $\theta$.
	We set the number of walkers and steps to 50 and $10^{4}$, respectively. 
	The initial positions of the parameters $h$, $w$, $\lambda$, and $\sigma$ 
	were set to [$10^{-3}$, $10^{-1}$], [$10^{-1}$, $10^{2}$], 
	[$10^{-1}$, $10^{2}$], and [$10^{-4}$, $10^{-2}$], respectively with uniform distributions.
	
	The derived hyperparameters in this analysis are listed in the upper part of Table \ref{tab:1}.
	We indicate the mean with a solid line and the $1 -\sigma$ and $2 -\sigma $ uncertainties 
	of the GP analyses with colored regions in Figure \ref{K2LC}.
	We see quasi-stable, periodic variations for all targets.
	
	\subsection{Joint GP Analysis}\label{ss:jointGP}
	\begin{figure*}[t]
 		\centering
 		\includegraphics[width=18cm]{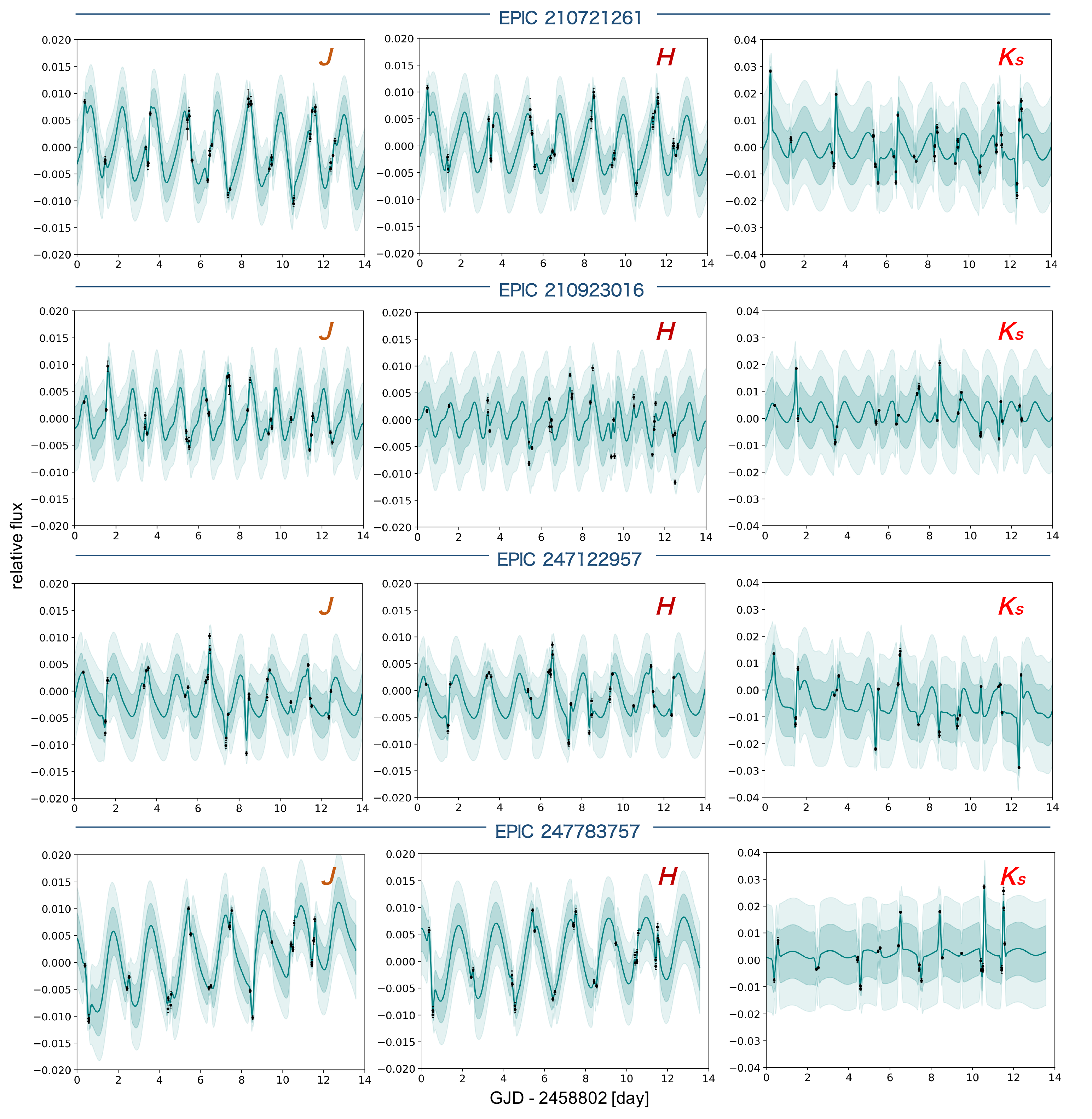}\\
 		\caption{\small Light curves for all observational bands for all targets plotted with black dots.
				The solid blue lines and dark-colored regions represent the mean
				and $1-\sigma$ variance results of the GP optimization, 
				with the light-colored regions representing the $2-\sigma$ variance results.
				From left to right the panels show $J$, $H$, and $K_s$.}%
 		\label{pic:resultGP}
	\end{figure*}
	Next, we measured the variation amplitudes $h$ over all the passbands ($Kp$, $J$, $H$, and $K_s$)
	using a Bayesian approach,
	to evaluate the wavelength dependencies of the stellar rotational activity.
	For the $K2$ light curves, 
	we applied the quasi-periodic kernel (Equation (\ref{kernel_qp})) 
	with $\theta$, $w$, and $\lambda$ 
	set to the values determined in Section \ref{GP_K2}, and re-derived $h$ and $\sigma$.
	We ran $10^4$ MCMC steps and ensured the convergence of the chains in the first $10^3$ steps.
	
	For the ground-based photometric data ($J$, $H$, and $K_s$), there are large systematic errors 
	arising from instrumental and/or weather conditions.
	Because of the lack of accurate modeling of these systematic errors,
	the astrophysical signals originating from the intrinsic stellar activity could be overestimated or underestimated.
	Using trial- and error- approach, we found that the following GP kernel, which combines 
	the quasi-periodic and squared exponential kernels can most effectively
	describe the behavior in the observed light curves:
	\begin{eqnarray}
		{\bf K}^{i, j} = {\bf K}_{\it qp}^{i, j} + {\bf K}_{\it sq}^{i},  \label{kernel_qpsp}
	\end{eqnarray}
	where the subscripts $i$ and $j$ indicate the passbands and the target ID, respectively.
	The first term ${\bf K}_{\it qp}^{i,j}$ is the quasi-periodic kernel 
	which reproduces the intrinsic stellar flux variations.
	We fixed the hyperparameters in ${\bf K}_{\it qp}^{i,j}$ to the values in the $Kp$ band except $h$ and $\sigma$.
	This is because the ground-based light curve data are sparse 
	and it is difficult to accurately derive all the relevant parameters for the stellar jitter from the ground-based photometry alone, 
	as seen in the treatment of the RV data in \citet[][]{Haywood2014}.
	Assuming that the covariance length scales and smoothing parameters are independent of the observational passband (wavelength), we allowed only $h$ and $\sigma$,
	which are expected to depend on the passband, 
	to float freely for each passband and each target.  
	The second term ${\bf K}_{\it sq}^{i}$ is the squared exponential kernel, 
	which only accounts for the instrumental systematic errors for each passband,
	in which the relevant hyperparameters are shared for all four targets.

	We list the mean values of each hyperparameter whose uncertainties were calculated 
	to be in the range of $68.3 \%$ from the median of the marginalized posterior distribution 
	in the lower part of Table \ref{tab:1}.
	The hyperparameters in ${\bf K}_{sq}^i$ corresponding to the correlated instrumental
	noise are depicted with additional subscripts ``$sq$".
	In Figure \ref{pic:resultGP}, 
	we also show the mean and $1-\sigma$ and $2-\sigma$ uncertainties of the GP regressions to the light curves
	for the $J$, $H$, and $K_s$ bands for each of the four targets.
	The smooth sinusoidal modulations represent the astrophysical signals modeled by ${\bf K}_{qp}^{i,j}$,
	whereas the 
	sudden fluctuations correspond to the systematic errors modeled by ${\bf K}_{sq}^{i}$.
	The latter variations act as offsets in the estimation of the signal amplitudes.
	In the $K_s$ band, 
	because the periodicities are weak for most of the targets as in Figure \ref{pic:lombscargle},
	the light curves are dominated by the sudden fluctuations due to instrumental effects. 
	In total, the estimated amplitudes in NIR are from approximately half to one-third of those in the $Kp$ band,
	while there appears to be no significant differences between the NIR passbands.
	In particular for EPIC 210721261, the amplitude ratios for the $Kp$ and $H$ bands are as large as 7:1.
	
	Generally, there is a possibility that the shape of the photometric variations changed 
	during $K2$ and IRSF observations, 
	even though we fixed the hyperparameter $w$ and $\lambda$ in this GP regression. 
	On the other hand, our four targets show stable photometric modulations as in Figure \ref{K2LC}.
	The periodogram analysis also suggests that they have single periodicities and 
	no significant differential rotation \citep{Reinhold2013}.
	\citet{Davenport2015} performed light curve analysis for a rapidly rotating ($\approx 0.6$ day) M-dwarf using $Kepler$ photometry
	and suggested that a starspot was very stable over many years.
	In addition, we succeeded to detect the rotation periods from the discrete ground-based photometry,
	which means the photometric variations were not complicated in the IRSF observations.
	Thus, we conclude that our estimations are likely consistent with the true nature of the targets.

\section{Estimated Starspot Properties}\label{sec:spots}
	\subsection{Starspot Variations over 2 Years}\label{ss:spotsvariation}
	\begin{figure}[]
 	\centering
 	\includegraphics[width=8cm]{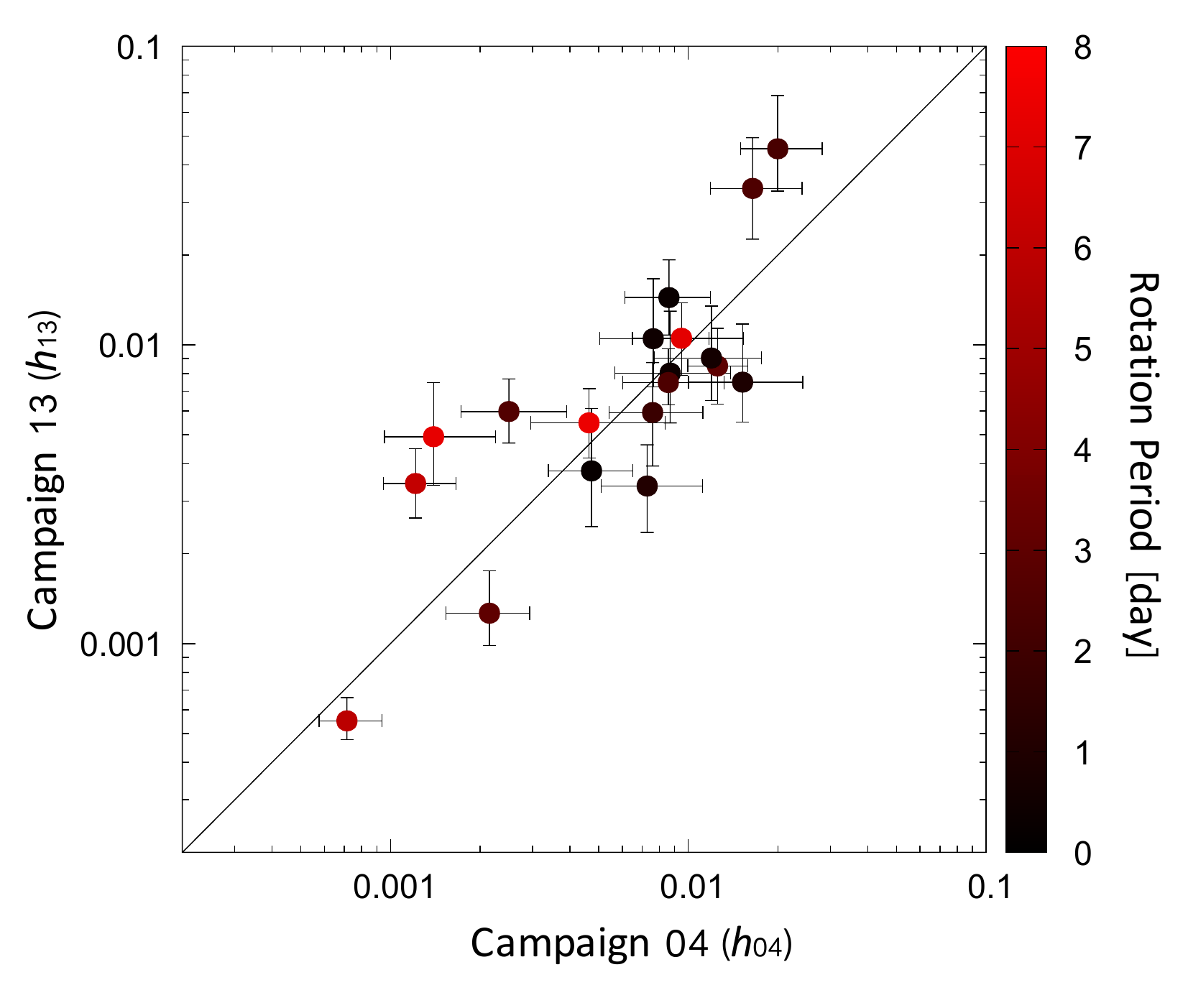}
	\centering
 	\caption{\small Comparison plot for the amplitudes of 19 M dwarfs in Hyades 
	between Campaigns 4 and 13 on a logarithmic scale.
	The color for each point represents a rotation period derived using the GLS periodogram.
	 The black solid line represents the $y = x$ line.}%
 	\label{pic:c04vsc13}
	\end{figure}
	%
	
	As noted in Section \ref{sec:targetIRSF}, the IRSF data were observed in November 2019,
	while the $K2$ data from Campaign 13 were observed from March 8 to May 27, 2017. 
	meaning that these two observations were not simultaneous but were separated by approximately  
	2.5 years. 
	Within this time interval, the properties of the surface activity (jitter) may have significantly altered.
	Therefore, before quantitatively comparing the flux-modulation amplitudes in the $K2$ and IRSF data, 
	we need to assess any long-term variations in the stellar surface activity. 
	To do so, we focused on multiple observations of the same stars in the $K2$ mission; 
	$K2$ observed the Hyades cluster during both Campaigns 4 and 13, 
	spanning a time interval of approximately 2 years. 
	By comparing the photometric data taken during the two different campaigns, 
	we can evaluate the long-term, temporal evolution of the surface activity. 
	Note that our targets which are rapidly rotating M-dwarfs are not typical in Hyades.
	Thus, we use targets whose rotation periods are less than 10 days 
	to focus on such a specific class of M-dwarfs \citep{Douglas2019}.
	
	While our targets were only observed in Campaign 13,
	we found that 333 EPIC targets were observed in both campaigns.
	Because we are focusing on cool stars in this study,
	we selected 95 targets with temperatures below $4000$ K 
	according the EPIC catalog \citep{Huber2016}.
	We normalized the light curves and removed the flux outliers as in Section \ref{sec:K2}.
	We computed the GLS periodograms to determine their rotation periods.
	The detection threshold in the GLS power for the $K2$ data was set to roughly 
	0.1 in reference to the periodograms in Figure \ref{pic:lombscargle}.
	Consequently, we detected 19 targets whose GLS peaks are larger than 0.1 
	and whose periods are shorter than 10 days in either of the two campaigns. 
	We measured the typical flux semi-amplitudes of the phase folded light curves 
	for the 19 targets for both campaigns using GP.
	Here, we use the periodic kernel ${\bf K}_{p}$ 
	in consideration of computational cost in the optimization with MCMC.
	\begin{figure*}[]
	\centering
 	\includegraphics[width=13cm]{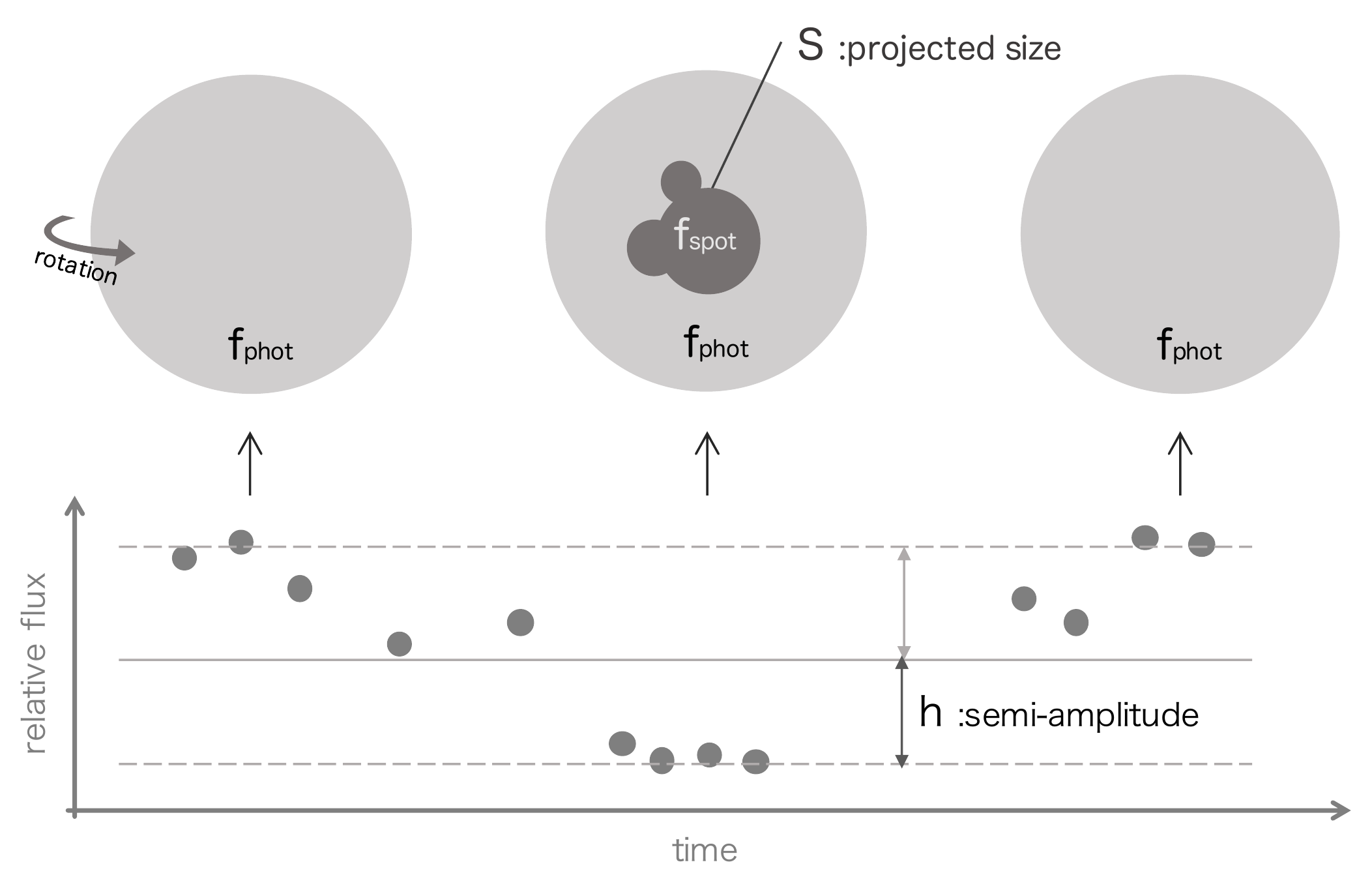}
 	\caption{\small Illustration of the simple model for the estimation of spot size and temperature (upper picture) and
	corresponding image of flux variation (lower picture),
	where $f_{\rm spot}$ and $f_{\rm phot}$ are fluxes on 
	the starspot and the photosphere regions, respectively;
	$h$ is semi-amplitude of the relative flux variation.
	The projected size of the starspot $S$ is defined to be 1.0 when covering a stellar hemisphere.
	Here, we do not consider the shape, surface distribution or rotation axis.
	}
 	\label{pic:model}
	\end{figure*}
	
	{Examples of the folded light curves and results of the GP regressions are shown 
	in Figure \ref{pic:MinHyades} in Appendix \ref{apb}.
	The properties and derived hyperparameters are listed in Table \ref{tab:19}.}
	One can see variations in the modulation amplitude and/or shape between the two campaigns.
	In Figure \ref{pic:c04vsc13}, we plot the results for all targets, showing the modulation amplitudes for Campaigns 4 and 13.
	The absolute variation in the amplitude is calculated as 
	$\Delta h = |h_{13} - h_{04}|$, 
	where $h_{04}$ and $h_{13}$ are the relative flux-variation amplitudes for Campaigns 4 and 13, respectively.
	We derived a weighted mean of $\Delta h$ for 19 targets of $0.0097 \pm 0.0094$.
	The relative variation with respect to Campaign 4 ($r \equiv \Delta h/h_{04}$) was determined to be $38\% \pm 71 \%$;
	therefore, that the variation in the modulation amplitudes at a timescale of $\sim 2$ years (between Campaigns 4 and 13) is likely lower than $100\%$ of the original flux modulation. 
	
	\begin{figure*}[]
	\centering
 	\includegraphics[width=15cm]{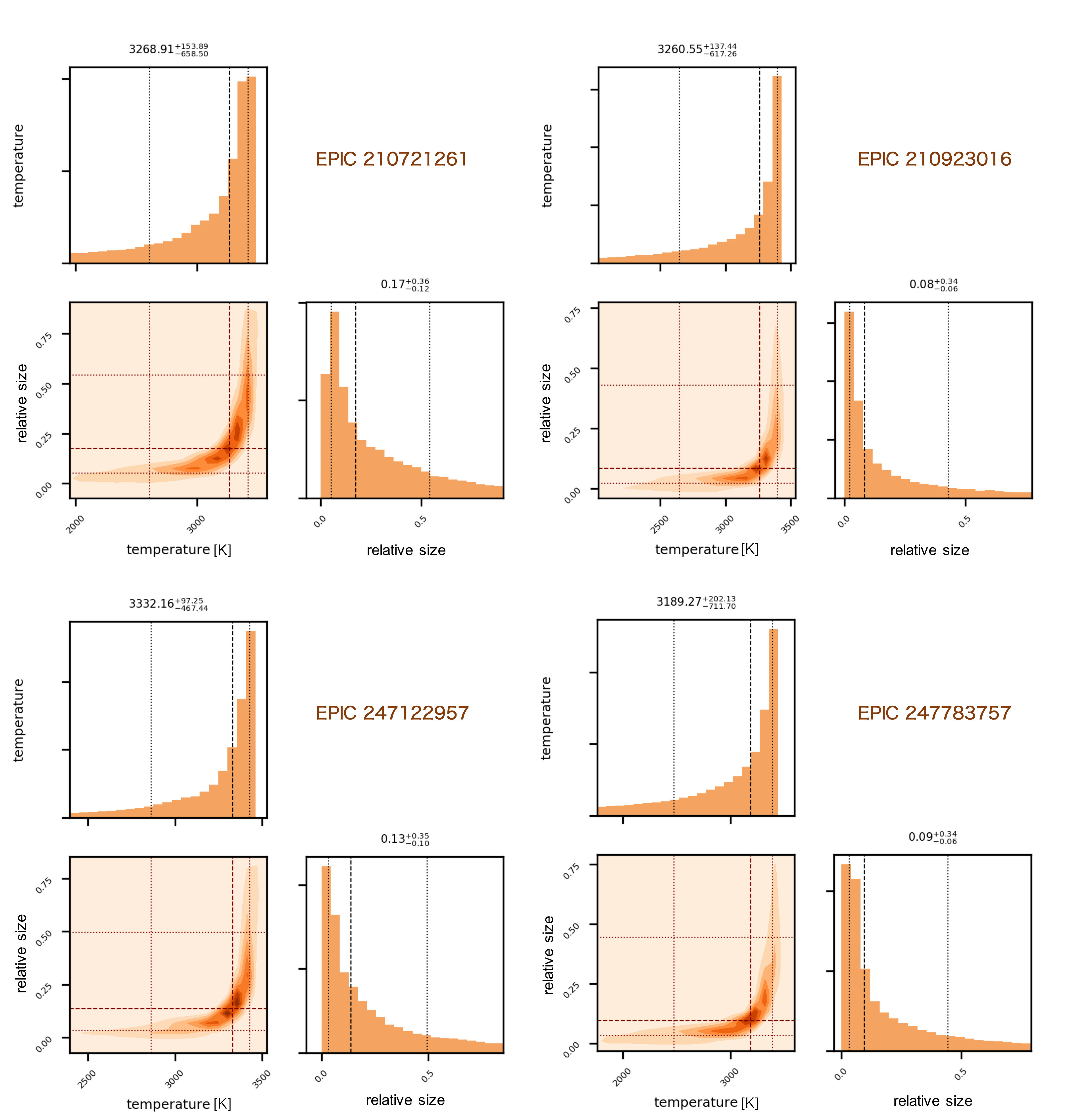}
 	\caption{\small Posterior distributions of Marcov Chain Monte Carlo (MCMC) sampling for estimations of the starspot temperatures and relative sizes.}
 	\label{pic:posterior}
	\end{figure*}

	\subsection{Estimations of Starspot Sizes and the Temperatures}
	\begin{figure*}[]
	\centering
 	\includegraphics[width=15cm]{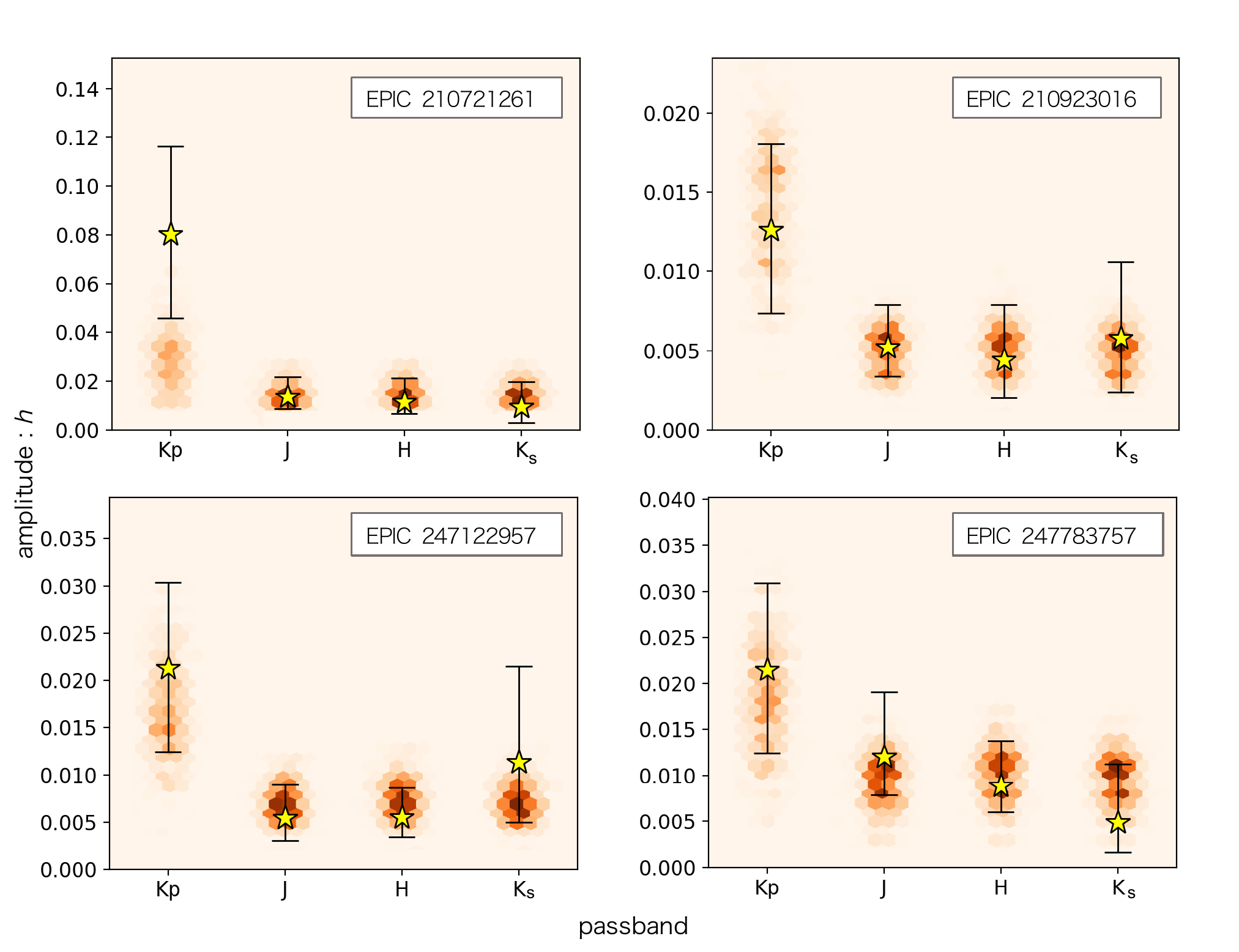}
 	\caption{\small Variation amplitude for each observational passband for our targets.
	The yellow stars represent the measured values in Section \ref{ss:jointGP}.
	The orange hexagons represent the model values derived from the posterior distribution
	of the MCMC sampling. {The error bars of the observed values for $Kp$ band were 
	enlarged by $r$, originating from the uncertainty in the spot evolution over the two years
	as in Section \ref{ss:spotsvariation}.}}
 	\label{pic:spot.pdf}
	\end{figure*}
	\begin{table*}[t]
	\centering
	\caption{\small Derived starspot sizes and temperatures}\label{tab:spot}
  		\begin{tabular}{ccccccccccc} \hline
    		\hline 
		EPIC ID 		&size [$\%$] 			& 	temperature [K] 			&  $T_{\rm spot}/T_{\rm phot}$
			&&size [$\%$]&$T_{\rm spot}/T_{\rm phot}$& &size[$\%$]& $T_{\rm spot}/T_{\rm phot}$ 
			\\ \hline  &&&&&\multicolumn{2}{c}{+200 K } &&\multicolumn{2}{c}{-200 K }
		\\ \cline{6-7} \cline{9-10}
		210721261 	& $18_{-12}^{+36}$ 	&	$3269_{-652}^{+153}$ 	&  $0.94_{-0.18}^{+0.04}$
			&&$13_{-9}^{+49}$&$0.94_{-0.22}^{+0.06}$&& $17_{-11}^{+35}$ & $0.93_{-0.17}^{+0.05}$\\
		210923016 	& $8_{-6}^{+34}$  	&	$3261_{-612}^{+137}$ 	&  $0.95_{-0.18}^{+0.04}$
			&&$5_{-3}^{+36}$&$0.93_{-0.22}^{+0.06}$& &$9_{-6}^{+36}$&$0.96_{-0.17}^{+0.04}$\\
		247122957 	& $14_{-10}^{+35}$ 	&	$3332_{-463}^{+97}$ 		&  $0.96_{-0.13}^{+0.03}$
			&&$8_{-5}^{+41}$&$0.96_{-0.20}^{+0.03}$ &&$11_{-8}^{+38}$& $0.95_{-0.14}^{+0.04}$\\
		247783757 	& $10_{-6}^{+34}$  	&	$3189_{-706}^{+202}$		&  $0.93_{-0.21}^{+0.06}$
			&&$6_{-4}^{+36}$&$0.91_{-0.23}^{+0.08}$& &$10_{-7}^{+34}$&$0.93_{-0.19}^{+0.05}$\\
		\hline
     		\end{tabular}
	\end{table*}
	To understand the behaviors and properties of starspots more quantitatively,
	we estimated their sizes and temperatures from the amplitudes measured in the multicolor photometry.
	In general, it is difficult to solve degeneracies on surface properties (e.g., starspot latitude, 
	number of starspots, and inclination) from shape of light curves 
	as described in \citep{Luger2021a}. 
	Therefore, we employed a very simple toy model, which considers 
	only the maximum projected size and temperature 
	(see Figure \ref{pic:model}),
	because we focus only on ``relative" variations between different passbands, 
	which are independent of the geometric structures of the photosphere and starspots.
	In the fiducial case, because the variation amplitudes are determined 
	from the appearance and disappearance of the starspots, 
	we ignore the effect of the limb-darkening in this analysis.
	The modeled semi-amplitude $h_{\rm model}$ is calculated as follows 
	\begin{eqnarray}
	h_{\rm model}	= (1 - f_{\rm spot}/f_{\rm phot})\times S/2,
	\end{eqnarray}
	where $f_{\rm spot }$ and $f_{\rm phot}$ are the fluxes of the starspot 
	and the photsphere, respectively,
	and $S$ represents the projected size of the starspot relative to the stellar disc,
	i.e., $S = 1.0$ means that a starspot covers a stellar hemisphere.
	We used the {\rm PHOENIX} atmosphere model \citep[\rm BT-SETTL ;][]{Allard2013} 
	to derive the photometric fluxes 
	from the temperatures of the spot ($T_{\rm spot}$) and photosphere ($T_{\rm phot}$).
	We generated the model spectra using a step of 100 K for 
	the effective temperature from 1600 K to 4000 K,
	and interpolated the intermediate values using a the third-order spline curve
	with the metallicity set to 0.0.
	We derived the photometric flux for each observational passband as the photon count 
	per unit area by multiplying the model spectra by the response function 
	for each passband 
	and integrating with respect to the wavelength \citep[e.g.,][]{Fukugita1995}.
	The response functions are taken from the websites for each instrument\footnote{https://keplerscience.arc.nasa.gov/data/kepler\_response\_hires1.txt}\footnote{http://www-ir.u.phys.nagoya-u.ac.jp/~irsf/sirius/tech/index.html}.
	We used the measured amplitudes $h$ in Section \ref{ss:jointGP} for the observed values.
	Because there are systematic uncertainties 
	due to the different observational windows for the $K2$ and IRSF runs, 
	as in Section \ref{ss:spotsvariation},
	we added the relative variation $r$ to the errors in the $K2$ data in quadrature such that 
	$({(h_{Kp} \times r)}^2 + \eta_{Kp}^2)^{1/2}$, 
	where $\eta_{Kp}$ is the internal error for $h_{Kp}$.
	The photospheric temperature $T_{\rm phot}$ was set to the stellar effective temperature 
	in Table \ref{tab:stellarparam}.
	For each target, we fitted the observed flux-modulation amplitude for each band by
	optimizing the projected size $S$ and the temperature $T_{\rm spot}$ of the starspot.  
	We ran $10^4$ MCMC steps with uniform priors in the range of [0.0, 1.0] and [1600, $T_{\rm phot}$) 
	for $S$ and $T_{\rm spot}$, respectively, 
	by maximizing the logarithmic likelihood $\log{\mathcal L}_h$ such that 
	\begin{eqnarray}
	\log{\mathcal L}_h \propto - \frac{1}{2}\sum_{i}{\frac{(h_{{\rm model}, i} - h_i )^2}{\eta_{i}^2}},
	\end{eqnarray}
	where $i$ is an index indicating the observational passband.
	
	The derived medians and $68.3 \%$ uncertainties of the posterior distributions 
	are given in Table \ref{tab:spot} and Figure \ref{pic:posterior}.
	We show the fitting results of the signal variations in Figure \ref{pic:spot.pdf}.
	The observed values are plotted with the yellow stars,
	and the posterior distributions of the modeled amplitudes $h_{\rm model}$ 
	are represented by orange hexagons 
	which are spread horizontally to easily discern each passband.
	Only in the case of EPIC 210721261 does the posterior distribution in the $Kp$ band 
	deviate significantly from the observed value.
	The other targets show good agreement for all passbands.
	The uncertainties with respect to the estimated sizes and temperatures are relatively large
	because the statistical error in $h$ derived from the photometry is large.
	For all targets, we can 
	see elongated posteriors in Figure \ref{pic:posterior} as a result of the degeneracy in 
	the starspot size and temperature. 
	
	To take into account the case that the photospheric temperatures were 
	mis-determined due to their young active natures,
	we performed additional analyses assuming $\pm 200$ K differences on the $T_{\rm phot}$.
	The results are also listed in Table \ref{tab:spot} 
	; there are no significant deviations from the fiducial values.
	Therefore, our conclusions on the $T_{\rm spot}$ are not likely severely affected 
	by the uncertainties in the photospheric temperature.
	We note that our modeling cannot solve degeneracies about the starspot size 
	if large polar spots and/or axis-symmetrically distributed spots exist on the photosphere,
	because they do not appear in the one-dimensional light curves.

\section{Discussion \& Summary}\label{sec:discussion}
	\begin{table}[t]
	\centering
	\caption{\small Estimated RV jitter from the photometry}\label{tab:jitter}
  		\begin{tabular}{crrrr} \hline
    		\hline 
		EPIC ID 		&$Kp\,({\rm ms}^{-1})$ 	& $J\,({\rm ms}^{-1})$ 	& $H \,({\rm ms}^{-1})$ & $K_s\,({\rm ms}^{-1})$ \\ 
		\hline
		210721261 	& 1370 &	230	&  	198 	& 160\\
		210923016 	&  259  &	106	& 	91	& 119\\
		247122957 	&  370	&	94 &  	95	& 196\\
		247783757 	&   280	&	156	&  	116	& 64\\
		\hline
     		\end{tabular}
	\end{table}
	
	\begin{figure*}[]
	\centering
 	\includegraphics[width=17cm]{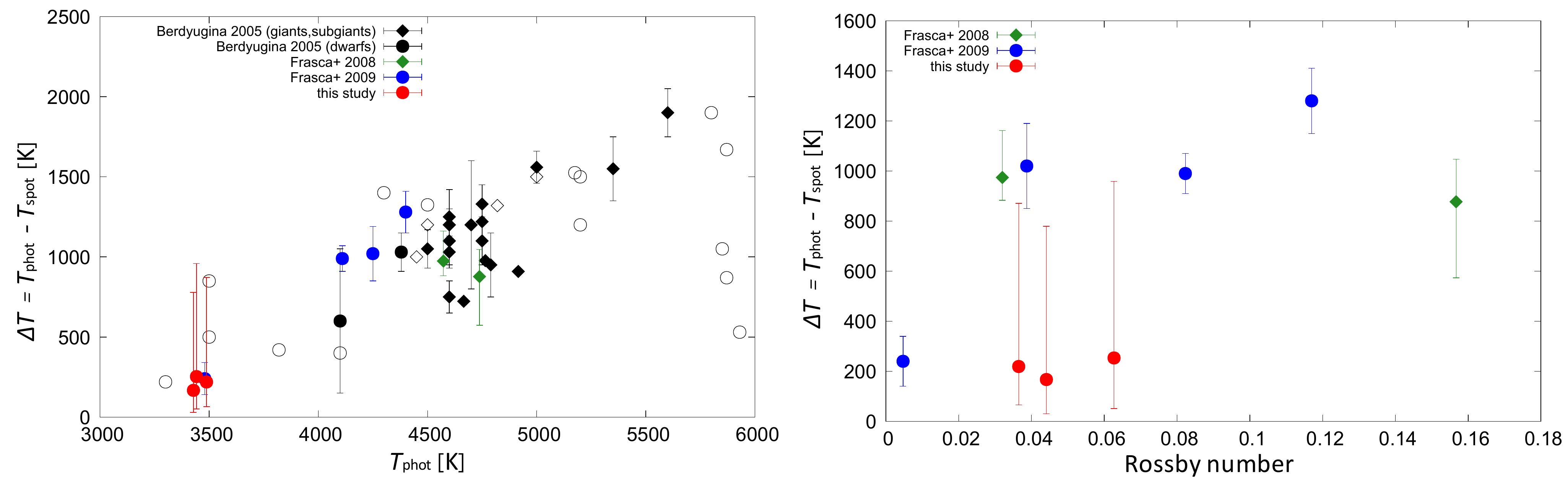}
 	\caption{\small Left: Temperature differences between the starspot and the photospere 
	with respect to the photospheric temperature.
	The black, green, blue and red points indicate values from \citet{Berdyugina2005}, \citet{Frasca2008},
	 \citet{Frasca2009} and this study,
	 respectively.
	The diamond and circular symbols represent giants/subgiants and dwarfs, respectively.
	We show the data points without errors with open circles from \citet{Berdyugina2005}.
	We omit EPIC 247122957 because of the high probability of binarity.
	Right: The temperature differences against the Rossby number.
	The same symbols and colors are used as in the left panel.
	}%
 	\label{pic:comparison}
	\end{figure*}
	\subsection{Wavelength Dependence of RV Jitter}
		We estimated the amplitudes of the flux modulations 
		due to stellar surface activity for the four targets in the Hyades cluster,
		and constrained the starspot sizes and temperatures using a simple toy model.
		For EPIC 210721261, the model in the $Kp$ band deviates from the observed amplitude,
		likely as a result of the particularly enhanced activity of the star 
		when the $K2$ observations were made. 
		EPIC 210721261 is the only target, 
		whose possibility of binarity is significantly low based on the works of \citet{Evans2018} with the Gaia DR2 data;
		for the other targets, it is possible that the starspot contrasts versus the wavelength
		are diluted by the presence of a companion.
		However, the Gaia EDR3 data \citep{Gaia2020} suggest binarity only for EPIC 247122957.
		In addition, because we assumed only cool starspots as elements of the surface activity,  
		as opposed to plages, flares, or hot starspots,
		our model may differ significantly from the true stellar photosphere.
		In either case, the reduction rate of the photometric variations 
		between the $Kp$ and $H$ bands ($1 - h_{H}/h_{Kp}$) is approximately 0.6 at maximum.
	
		To predict RV jitter in the observing bandpasses,
		we approximated the maximum amplitudes as $h \times V_{\rm eq}$ \citep{Aigrain2012}
		when the starspot is on the equator and the rotational inclination is $90^{\circ}$, 
		where $V_{\rm eq}$ is the equatorial rotational velocity; 
		these approximations are listed in Table \ref{tab:jitter}.
		Consequently, we found that RV jitter in the NIR is significantly suppressed 
		compared to jitter in visible wavelengths 
		and that there is no significant difference between the $J$, $H$, and $K_s$ passbands.
	
	\begin{figure*}[t]
	\centering
 	\includegraphics[width=16cm]{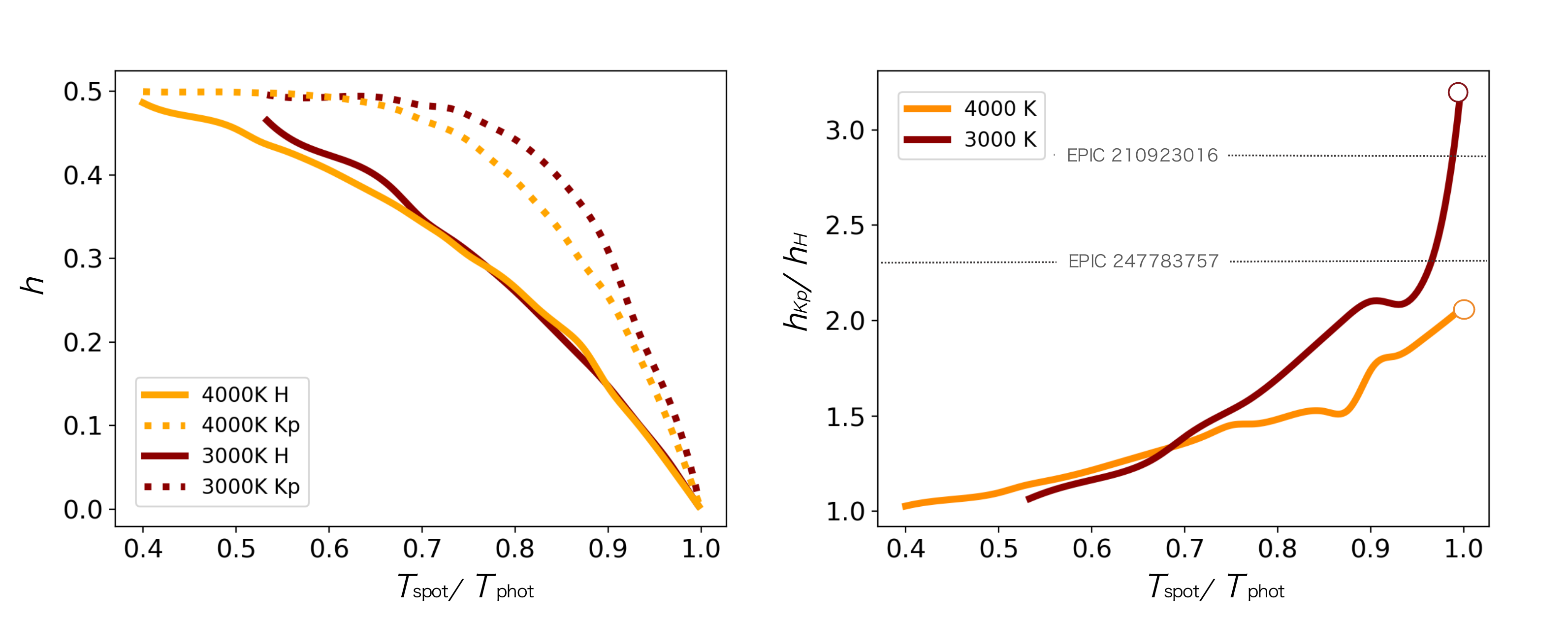}
 	\caption{\small Left: Theoretical relationships between the relative starspot temperature
	and the relative flux semi-amplitude $h$ when the projected size $S$ is 1.0  
	calculated with the PHOENIX model spectra.
	The solid and the dotted lines to $H$ band and $Kp$ band, and 
	the orange and the red lines to stellar photospheric temperatures 
	of 4000 K and 3000 K, respectively.
	Right: Theoretical relationship between the relative starspot temperature 
	and the ratio of the variation of the $H$ band with respect to the  $Kp$ band.
	The line color relationship is the same as the left figure.
	The gray horizontal lines indicate the measured values 
	for EPIC 210923016 and EPIC 247783757 as a reference;
	note that the values for the other two targets are beyond the figure frame.
	}%
	\label{pic:contrast.pdf}
	\end{figure*}

	\subsection{Comparison with Previous Studies of the Starspot Property}

	\citet{Frasca2009} measured the starspot properties of six cool stars 
	whose photometric variations are typically $\Delta {\rm mag}_H\approx0.1$ ($\sim 10 \%$ in relative flux)
	in the pre-main-sequence phase  ($\sim 10 $ Myr) 
	using the simultaneous multicolor photometry in the $R$, $I$, $J$, and $H$ bands.
	They derived the starspot size $S$ and the temperature ratio between the starspot 
	and the photosphere $T_{\rm spot}/T_{\rm phot}$
	to be 5$\%$ - 10$\%$ and 0.70 - 0.90, respectively. 
	Our estimations are 8$\%$- 18$\%$ and $0.93 - 0.96$, respectively, for 650-Myr cool stars.
	Even though the sizes depend on the selection biases of the targets,
	these results may suggest that the temperature difference 
	between starspot and the photosphere becomes smaller with stellar age.
	\begin{figure}[]
	\centering
 	\includegraphics[width=8cm]{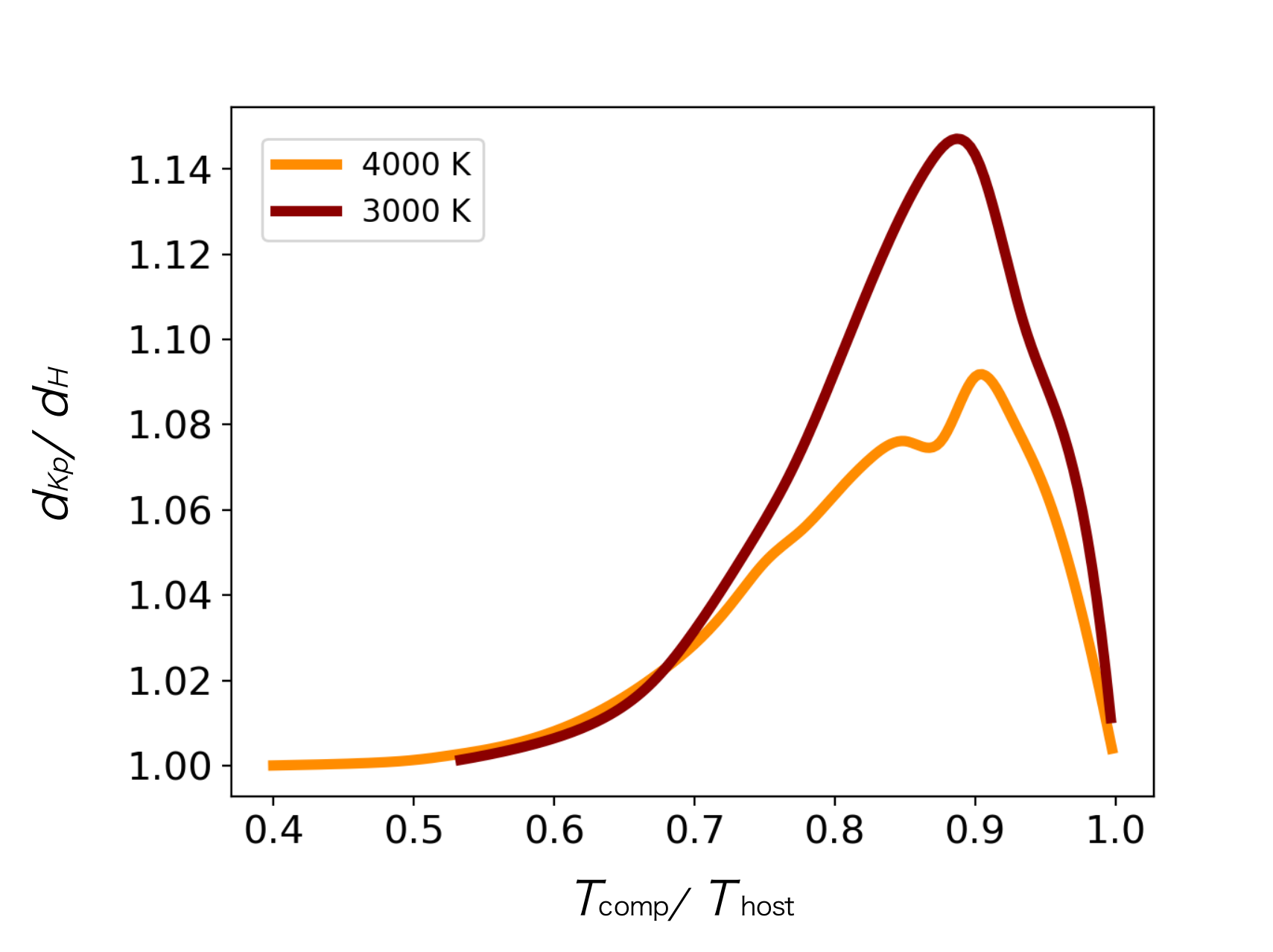}
 	\caption{\small Theoretical relationship between the relative companion temperature 
	and the ratio of the variation of the $H$ band with respect to the $Kp$ band. 
	The line color relationship is the same as Figure \ref{pic:contrast.pdf}.}%
 	\label{pic:companion}
	\end{figure}

	In some previous studies, the starspot to photosphere temperature ratio $T_{\rm spot}/T_{\rm phot}$ 
	was estimated to be approximately $0.8$ 
	for G and K -type stars according to line-depth ratio measurements, 
	which are currently believed to be the most reliable method to characterize the properties of starspots
	\citep[e.g.,][]{Catalano2002, Frasca2008}.
	In addition, \citet{ONeal1996} and \citet{Frasca2008} suggested 
	that $\Delta T = T_{\rm spot} - T_{\rm phot}$ increases with surface gravity of the star,
	which may be explained by the balance of magnetic and gas pressures in flux tubes.
	However, our estimation appears to be systematically inconsistent 
	with their theory considering the high surface gravity of our targets in Table \ref{tab:stellarparam}.
	In addition, from results found in different models, \citet{Berdyugina2005} 
	and \citet{ Strassmeier2009} showed that $\Delta T$ gets smaller for cooler stars 
	by as much as 200 K.
	We plot the previous $\Delta T$ estimations from \citet{Berdyugina2005, Frasca2008, Frasca2009} 
	in the left panel of Figure \ref{pic:comparison}. 
	Our results with the red points seem to follow this trend, 
	although the derived values of $\Delta T$ are estimated to be slightly small.
	The 3D radiative MHD simulations of starspots performed in \citet{Panja2020} also explained this trend
	with the dependence of opacity on temperature which is largely governed with $\rm H^{-}$ ions.
	Note that previous estimations of M dwarfs were derived using single-band 
	light curve modeling \citep[e.g., ][]{Rodono1986}, 
	which includes degeneracy between the starspot temperature and size.
	In addition, the samples used include large systematic uncertainties associated with the stellar evolution stage.
	In the right panel of Figure \ref{pic:comparison}, we also show $\Delta T$ with respect to the Rossby number which is 
	related to the magnetic activity in stellar dynamo \citep{Kim1996} 
	and is approximated as $Ro = v/(2\Omega L)$; 
	$v$, $\Omega$, and $L$ are surface velocity, angular velocity, and typical length, respectively.
	Here, we use the square root of the spot area for $L$, 
	and macro-turbulence velocity for $v$, 
	which is calculated assuming that $v$ linearly depends on the photospheric temperature
	in the range of $3000 < T_{\rm phot} < 4500$ K and $1 < v < 2$ kms$^{-1}$, respectively.
	$\Delta T$ for our targets are significantly lower than the previous estimations 
	for pre-main-sequence stars
	and the systematic trend with $Ro$ is unclear.
	In any case, additional investigations are required for further relevant discussions.
	
	\subsection{Model Uncertainty}
	In Figure \ref{pic:contrast.pdf}, 
	we plot the theoretical semi-amplitude ($h_{Kp}$ and $h_{H}$) with setting $S$ to 1.0 in the left panel 
	and the corresponding contrast ($h_{Kp}/h_{H}$) in the right panel 
	as a function of the temperature ratio $T_{\rm spot}/ T_{\rm phot}$ 
	in cases where the photospheric temperature is either 4000 K or 3000 K
	with the PHOENIX model spectra.
	The gray lines represent the values for EPIC 210923016 (3428 K) and EPIC 247783757 (3443 K), respectively,  
	whose contrasts are relatively small in our targets.
	This contrast figure suggests that 
	as the starspot temperature asymptotically approaches the surface temperature, 
	the contrasts between the starspot and the photosphere for the optical and NIR passbands become larger.
	Therefore, the starspots need to be hot at the same level as the photosphere 
	to explain our observational results.
	Nevertheless, the observed contrasts are still larger than the theoretical model expectations for EPIC 210721261, EPIC 210923016, and EPIC 247122957.
	Note that contrasts for EPIC 210721261 and EPIC 247122957 are beyond the range of the figure.
	This discrepancy is preferable for the detection of true planetary signals via NIR
	observations, albeit the reason for this may stem from the incompleteness of the models.
	
	We also estimated the amplitude variation ratio between the $H$ and $Kp$ bands 
	due to an unresolved cool companion.
	We calculated fluxes ($f_{\rm host}$, $f_{\rm comp}$) with the PHOENIX model spectra
	and derived the radii ($r_{\rm host}$, $r_{\rm comp}$) with a temperature-radius relationship in \citet{Mann2015} for the host and the companion, respectively.
	The signal dilution $d$ due to the companion is derived as,
	\begin{eqnarray}
		d 	= f_{\rm host}/ \Biggl\{f_{\rm host} + f_{\rm comp}\biggl(\frac{r_{\rm comp}}{r_{\rm host}}\biggr)^2\Biggr\} .\nonumber
	\end{eqnarray}
	The contrast between the bands is estimated as $d_{Kp}/d_{H}$
	and shown in Figure \ref{pic:companion}.
	The maximum variation is approximately $15 \%$
	which is significantly smaller than the variation due to the starspot.
	For example, the starspot temperature would be $3330_{-492}^{+100}$ K for EPIC 247122957,
	even after the consideration of the $15 \%$ amplitude decrease in $Kp$ band.
	Therefore, our conclusions are not affected seriously even though the targets are binaries.
	
	
	
	\subsection{Advantage of This Study}
	Previous studies on starspots were performed for bright targets 
	whose high-resolution spectra are available \citep{Catalano2002}
	or whose photometric variations are sufficiently large ($\sim 10 \%$), 
	such as pre-main-sequence stars \citep{Frasca2009}.
	Our approach, which combines space telescope and ground-based photometry, 
	is applicable to a larger sample of targets with relatively small variations ($\sim 1 \%$).
	Even though the simple modeling to the multicolor photometry still includes degeneracies on geometry
	of the starspots,
	it is useful for investigating the macro temperature structure on the stellar surface.
	Because the $K2$ and $TESS$ missions collected and are collecting many light curves 
	of young stars belonging to various stellar clusters, 
	a similar approach to that presented here could reveal  
	the statistical properties of stellar activities at different ages.
	
	Finally, we suggest that the mitigation of RV jitter in the NIR could be more significant than expected from theoretical models, 
	even though further observations with a larger sample are required to corroborate this possibility. 
	To take advantage of NIR spectroscopy in Doppler observations,  
	new NIR high-resolution spectrographs have been developed over the last decade, 
	such as CARMENES at the Calar Alto 3.5-m telescope \citep[][]{Quirrenbach2016},
	the Habitable-Zone Planet Finder \citep[][]{Mahadevan2014} on the Hobby Eberly Telescope,
	SPIRou on the CFHT 3.58-m telescope \citep[][]{Artigau2014},
	and the InfraRed Doppler spectrograph on the Subaru 8.2-m 
	\citep[][]{Tamura2012, Kotani2014,Kotani2018}.
	Our results support the effectiveness of these NIR observations.
	In the near future, more accurate estimations of the properties of planetary systems around young stars will be possible by combining photometric and spectroscopic NIR observations,
	which will offer important clues understanding the formation and evolution processes
	of exoplanetary systems. 
	
\acknowledgments
	
	This paper is based on data collected at IRSF.
	We thank Kumiko Morihana and the other members of IRSF team for support of the observation. 
	This work was supported by Japan Society for Promotion of Science (JSPS) KAKENHI Grant Numbers JP19J21733 and 17H04574.
 	This work has made use of data from the European Space Agency (ESA) mission
	Gaia (\url{https://www.cosmos.esa.int/gaia}), processed by the Gaia
	Data Processing and Analysis Consortium (DPAC,
	\url{https://www.cosmos.esa.int/web/gaia/dpac/consortium}). Funding for the DPAC
	has been provided by national institutions, in particular the institutions
	participating in the Gaia Multilateral Agreement.
	
	\appendix
	\section{GP Regression}\label{apa}
	Gaussian Process (GP) is a non-parametric regression technique 
	to analyze observed data having {\it n} points.
	It models an $n\,\times\,n$ covariance matrix, that expresses the correlation 
	between the data points using kernel functions.
	In this study, we used the ``{\it squared exponential}" kernel, ``{\it periodic}" kernel 
	and ``{\it quasi-periodic}" kernel, as used in, e.g., \citet{Grunblatt2015} as follows:
	\begin{eqnarray}
	 	{\bf K}_{\it sq} & ~=~ & h^2 \exp ~\Biggl[ ~-~ \Bigl( \frac{t_i - t_j}{\lambda}\Bigr)^2 \Biggr],
		  \label{kernel_sq} \\ \nonumber \\	
		  \nonumber \\ 
		{\bf K}_{\it p} & ~=~ & h^2 \exp ~\Biggl[ ~-~ \frac{\sin^2\{\pi (t_i - t_j)/\theta\}}{2w^2} \Biggr] ,
		 \label{kernel_p}\\  \nonumber \\
		  \nonumber \\ 
	 	{\bf K}_{\it qp} & ~=~ & h^2 \exp ~\Biggl[ ~-~ \frac{\sin^2\{\pi (t_i - t_j)/\theta\}}{2w^2}
		 	~-~ \Bigl( \frac{t_i - t_j}{\lambda}\Bigr)^2 \Biggr] ,\label{kernel_qp}\\  \nonumber
	\end{eqnarray}
	where $t$ is the data point of time, $h$ is the covariance amplitude, 
	$\lambda$ is the covariance length scale,  $\theta$ is the period of the variation, 
	and $w$ is the smoothing parameter of 
	the periodicity for each kernel. 
	The squared-exponential kernel expressed by Equation (\ref{kernel_sq}) 
	reproduces the continuous data points in the observed signals,
	and is often used for estimations and/or corrections to systematic errors.
	Equation (\ref{kernel_p}) is the periodic kernel, which is used to flexibly model periodic signals
	including non-sinusoidal variations.
	The quasi-periodic kernel consisting of periodic 
	and squared-exponential components (Equation (\ref{kernel_qp}))
	is often used to model quasi-periodic signals, including coherent modes
	such as stellar jitter.
	
	We optimized the hyperparameters of the kernels by maximizing the likelihood ${\mathcal L}$.
	Under the assumption that ${\mathcal L}$ follows an $n$-dimensional Gaussian distribution, 
	the logarithmic likelihood is described with observed data points $y$ as
	\begin{eqnarray}
		\log {\mathcal L} = -~\frac{n}{2}\log (2\pi) ~-~ \frac{1}{2}\log (|{\bf K} + \sigma^2{\bf I}|) \nonumber \\ 
		- ~\frac{1}{2} \underline{y}^{T}\cdot ({\bf K} + \sigma^2{\bf I})^{-1} \cdot \underline{y}, \nonumber
	\end{eqnarray}
	where ${\bf I}$ represents the identity matrix,
	$\underline{y}$ is the vector of the residuals of $y$ from the mean, and
	$\sigma$ is white noise, which represents the statistical uncertainty of each data point.
	In the maximization, we optimized the white noise component 
	by including the internal uncertainty of the observed data point ($\sigma_{\rm internal}$; i.e., photon noise) 
	as $\sigma = (\sigma_{\rm internal}^2 + \sigma_{\rm white}^2)^{1/2}$.
	We performed the MCMC analysis provided 
	by the Python package \texttt{emcee} \citep{Foreman-Mackey2013}
	to maximize the likelihood. 
	
	\section{Summary of the 19 targets in K2}\label{apb}
\begin{figure*}[]
 	\centering
 	\includegraphics[width=19cm]{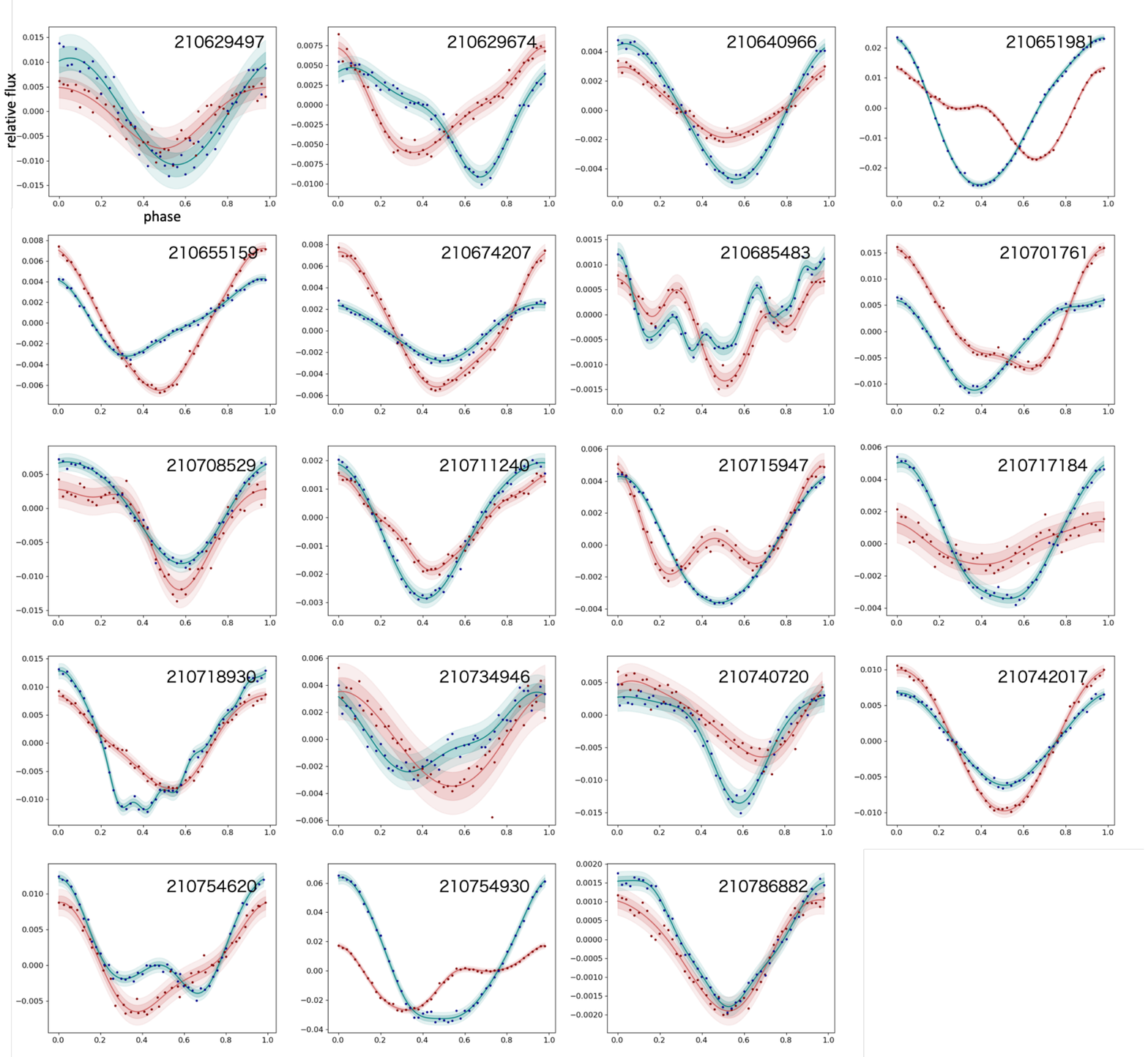}
 	\caption{\small {Examples of phase-folded light curve for 19 Hyades M dwarfs.}
	The red and blue data are in Campaign 04 and 13, respectively.
	The solid lines and colored regions represent the means and variances derived using GP 
	with a periodic kernel.}
	\label{pic:MinHyades}
	\end{figure*}

	Here, we summarize the properties of the 19 targets in Section \ref{ss:spotsvariation}.
	The effective temperature derived in \citet{Huber2016}, the rotation period with GLS 
	\citep{Zechmeister2009}, and the hyperparameters of the periodic kernel in GP 
	are listed in Table \ref{tab:19}.
	{The results of the GP analyses for all the targets in both Campaign 04 \& 13
	are shown in Figure \ref{pic:MinHyades}.}
	
	\begin{table}[]
	\centering
	\caption{\small Derived properties of 19 targets}\label{tab:19}
  		\begin{tabular}{cccccccccccc} \hline
    		\hline 
		EPIC ID & Temperature [K] (1) & Period [day] (2)& & $h_{04}$& $w_{04}$ & $\sigma_{04}$ && $h_{13}$& $w_{13}$ & $\sigma_{13}$ \\ \hline
		210629497 &  	3567&	0.27&&	$0.0086_{-0.0024}^{+0.0032}$	&$2.38_{-0.59}^{+0.43}$	&$0.0020_{-0.0001}^{+0.0001}$&
		&	$0.0143_{-0.0036}^{+0.0048}$	&$2.55_{-0.52}^{+0.32}$	&$0.0023_{-0.0001}^{+0.0001}$\\
		210629674 &  	3523&	0.40&&	$0.0086_{-0.0030}^{+0.0051}$	&$1.82_{-0.44}^{+0.56}$	&$0.0007_{-0.0000}^{+0.0000}$&
		&	$0.0080_{-0.0025}^{+0.0049}$	&$1.58_{-0.34}^{+0.47}$	&$0.0006_{-0.0000}^{+0.0000}$\\
		210640966 &  	3387&	2.55&&	$0.0024_{-0.0007}^{+0.0014}$	&$1.57_{-0.40}^{+0.58}$	&$0.0003_{-0.0000}^{+0.0000}$&
		&	$0.0059_{-0.0012}^{+0.0017}$	&$2.70_{-0.38}^{+0.21}$	&$0.0003_{-0.0000}^{+0.0000}$\\
		210651981 &  	3712&	2.44&&	$0.0164_{-0.0045}^{+0.0076}$	&$1.42_{-0.19}^{+0.24}$	&$0.0004_{-0.0000}^{+0.0000}$&
		&	$0.0333_{-0.0107}^{+0.0160}$	&$2.15_{-0.43}^{+0.43}$	&$0.0005_{-0.0000}^{+0.0000}$\\
		210655159 &  	3805&	1.83&&	$0.0075_{-0.0021}^{+0.0036}$	&$1.87_{-0.36}^{+0.49}$	&$0.0002_{-0.0000}^{+0.0000}$&
		&	$0.0059_{-0.0020}^{+0.0027}$	&$2.21_{-0.49}^{+0.48}$	&$0.0002_{-0.0000}^{+0.0000}$\\
		210674207 &  	3071&	1.05&&	$0.0072_{-0.0021}^{+0.0038}$	&$1.76_{-0.32}^{+0.42}$	&$0.0004_{-0.0000}^{+0.0000}$&
		&	$0.0033_{-0.0010}^{+0.0012}$	&$2.43_{-0.80}^{+0.40}$	&$0.0002_{-0.0000}^{+0.0000}$\\
		210685483 &  	3800&	5.86&&	$0.0007_{-0.0001}^{+0.0002}$	&$0.63_{-0.08}^{+0.10}$	&$0.0001_{-0.0000}^{+0.0000}$&
		&	$0.0005_{-0.0000}^{+0.0001}$	&$0.34_{-0.04}^{+0.08}$	&$0.0001_{-0.0000}^{+0.0000}$\\
		210701761 &  	3451&	0.88&&	$0.0152_{-0.0052}^{+0.0089}$	&$1.82_{-0.42}^{+0.50}$	&$0.0005_{-0.0000}^{+0.0000}$&
		&	$0.0075_{-0.0019}^{+0.0042}$	&$1.21_{-0.25}^{+0.61}$	&$0.0005_{-0.0000}^{+0.0000}$\\
		210708529 &  	3538&	7.16&&	$0.0094_{-0.0030}^{+0.0058}$	&$1.49_{-0.32}^{+0.46}$	&$0.0012_{-0.0000}^{+0.0001}$&
		&	$0.0105_{-0.0026}^{+0.0032}$	&$2.59_{-0.45}^{+0.28}$	&$0.0006_{-0.0000}^{+0.0000}$\\
		210711240 &  	3500&	6.11&&	$0.0012_{-0.0002}^{+0.0004}$	&$0.99_{-0.12}^{+0.15}$	&$0.0001_{-0.0000}^{+0.0000}$&
		&	$0.0034_{-0.0008}^{+0.0010}$	&$2.64_{-0.42}^{+0.25}$	&$0.0001_{-0.0000}^{+0.0000}$\\
		210715947 &  	4000&	7.39&&	$0.0046_{-0.0016}^{+0.0037}$	&$1.41_{-0.34}^{+0.52}$	&$0.0003_{-0.0000}^{+0.0000}$&
		&	$0.0054_{-0.0013}^{+0.0016}$	&$2.63_{-0.41}^{+0.26}$	&$0.0001_{-0.0000}^{+0.0000}$\\
		210717184 &  	3679&	7.24&&	$0.0013_{-0.0004}^{+0.0008}$	&$1.44_{-0.68}^{+0.98}$	&$0.0005_{-0.0000}^{+0.0000}$&
		&	$0.0049_{-0.0015}^{+0.0025}$	&$1.81_{-0.47}^{+0.62}$	&$0.0002_{-0.0000}^{+0.0000}$\\
		210718930 &  	3504&	2.41&&	$0.0085_{-0.0025}^{+0.0046}$	&$1.57_{-0.35}^{+0.64}$	&$0.0006_{-0.0000}^{+0.0000}$&
		&	$0.0074_{-0.0011}^{+0.0022}$	&$0.46_{-0.03}^{+0.53}$	&$0.0005_{-0.0000}^{+0.0002}$\\
		210734946 &  	3550&	0.20&&	$0.0047_{-0.0013}^{+0.0017}$	&$2.39_{-0.61}^{+0.43}$	&$0.0009_{-0.0000}^{+0.0000}$&
		&	$0.0037_{-0.0013}^{+0.0023}$	&$1.77_{-0.46}^{+0.62}$	&$0.0006_{-0.0000}^{+0.0000}$\\
		210740720 &  	3625&	0.46&&	$0.0076_{-0.0025}^{+0.0041}$	&$1.90_{-0.51}^{+0.62}$	&$0.0012_{-0.0000}^{+0.0001}$&
		&	$0.0104_{-0.0032}^{+0.0061}$	&$1.54_{-0.29}^{+0.41}$	&$0.0010_{-0.0000}^{+0.0000}$\\
		210742017 &  	3524&	2.88&&	$0.0125_{-0.0025}^{+0.0033}$	&$2.75_{-0.33}^{+0.18}$	&$0.0004_{-0.0000}^{+0.0000}$&
		&	$0.0085_{-0.0021}^{+0.0028}$	&$2.52_{-0.46}^{+0.33}$	&$0.0003_{-0.0000}^{+0.0000}$\\
		210754620 &  	3512&	0.63&&	$0.0119_{-0.0042}^{+0.0056}$	&$2.18_{-0.54}^{+0.51}$	&$0.0008_{-0.0000}^{+0.0000}$&
		&	$0.0090_{-0.0025}^{+0.0044}$	&$1.32_{-0.20}^{+0.28}$	&$0.0004_{-0.0000}^{+0.0000}$\\
		210754930 &  	3836&	2.26&&	$0.0199_{-0.0049}^{+0.0081}$	&$1.16_{-0.14}^{+0.18}$	&$0.0009_{-0.0000}^{+0.0000}$&
		&	$0.0453_{-0.0125}^{+0.0229}$	&$1.40_{-0.30}^{+0.61}$	&$0.0020_{-0.0001}^{+0.0002}$\\
		210786882 &  	3700&	2.99&&	$0.0021_{-0.0006}^{+0.0007}$	&$2.45_{-0.56}^{+0.38}$	&$0.0001_{-0.0000}^{+0.0000}$&
		&	$0.0012_{-0.0002}^{+0.0004}$	&$1.02_{-0.14}^{+0.19}$	&$0.0001_{-0.0000}^{+0.0000}$
 \\ \hline

		\hline
     		\end{tabular}
	\begin{flushleft}
  			{\bf References :} (1): \citet{Huber2016}, (2) GLS in this study.
  	 	\end{flushleft}
	\end{table}
	
 \bibliographystyle{aasjournal}
 \bibliography{miyakawa20.bib}

\end{document}